\documentstyle[11pt,paspconf,psfig]{article}
\def\cor{\mathrel{\mathchoice {\hbox{$\widehat=$}}{\hbox{$\widehat=$}}
{\hbox{$\scriptstyle\hat=$}}
{\hbox{$\scriptscriptstyle\hat=$}}}}
\def\la{\mathrel{\mathchoice {\vcenter{\offinterlineskip\halign{\hfil
 $\displaystyle##$\hfil\cr<\cr\sim\cr}}}
 {\vcenter{\offinterlineskip\halign{\hfil$\textstyle##$\hfil\cr
 <\cr\sim\cr}}}
 {\vcenter{\offinterlineskip\halign{\hfil$\scriptstyle##$\hfil\cr
 <\cr\sim\cr}}}
 {\vcenter{\offinterlineskip\halign{\hfil$\scriptscriptstyle##$\hfil\cr
 <\cr\sim\cr}}}}}
\def\ga{\mathrel{\mathchoice {\vcenter{\offinterlineskip\halign{\hfil
 $\displaystyle##$\hfil\cr>\cr\sim\cr}}}
 {\vcenter{\offinterlineskip\halign{\hfil$\textstyle##$\hfil\cr
 >\cr\sim\cr}}}
 {\vcenter{\offinterlineskip\halign{\hfil$\scriptstyle##$\hfil\cr
 >\cr\sim\cr}}}
 {\vcenter{\offinterlineskip\halign{\hfil$\scriptscriptstyle##$\hfil\cr
 >\cr\sim\cr}}}}}

\newcommand{\Lsun}{\,\mbox{L}_{\odot}}

\newcommand{\scrm}[1]{\mbox{\scriptsize\rm #1}}
\newcommand{\scsl}[1]{\mbox{\scriptsize\sl #1}}
\newcommand{\scsc}[1]{\mbox{\scriptsize\sc #1}}
\newcommand{\tinm}[1]{\mbox{\tiny\rm #1}}

\newcommand{\etal}{{\it et al.\/}~}

\newcommand{\ie}{{\it i.e.},\ }
\newcommand{\eg}{{\it e.g.},\ }

\newcommand{\Mzon}{M$_{\odot}$}
\newcommand{\Lzon}{L$_{\odot}$}

\newcommand{\EBV}{$E_{B-V}$}

\newcommand{\kms}{\mbox{km s$^{-1}$}}

\newcommand{\Mpc}{Mpc$^{-1}$} 
\newcommand{\Ha}{{\sc H$\alpha$}}
\newcommand{\Hb}{{\sc H$\beta$}}
\newcommand{\lda}{$\lambda$}

\newcommand{\NeIII}{{\rm [Ne$\,${\sc iii}]}}
\newcommand{\NII}{{\sc [N$\,$ii]}}
\newcommand{\OII}{{\sc [O$\,$ii]}}
\newcommand{\OIII}{{\sc [O$\,$iii]}}
\newcommand{\SII}{{\sc [S$\,$ii]}}

\newcommand{\OI}{{\sc [O$\,$i]}}
\newcommand{\HI}{{\sc H$\,$i}}
\newcommand{\HII}{{\sc H$\,$ii}}

\begin{document}
 
\title{The Nature of Ionized Gas in Giant Elliptical Galaxies} 

\author{Paul Goudfrooij\,\footnote{Affiliated to the Astrophysics
Division, Space Science Department, European Space Agency}}
%\altaffilmark{1}}
\affil{Space Telescope Science Institute \\ 3700 San Martin Drive,
Baltimore, MD 21218, U.S.A.}

\altaffiltext{1}{Affiliated to the Astrophysics Division, Space
Science Department, European Space Agency}
% ESA, ESTEC, Postbus 299, NL--2200 AG Noordwijk, The Netherlands

%\addtocounter{footnote}{1}
 
\begin{abstract}
I review the current understanding of the origin and nature of
``warm'' ionized gas in giant elliptical galaxies in the light of
results of recent imaging and spectroscopic surveys. CCD imaging
surveys have revealed that emission-line disks or filaments are (as a
rule) associated with dust absorption, even in the X-ray brightest
systems. This strongly suggests that the origin of this ionized gas is
generally not through ``cooling flows''; galaxy interactions are 
favored instead. 
Using data from a new spectrophotometric survey of ``normal''
elliptical galaxies covering the whole optical range, the extended
ionized gas in giant ellipticals 
is found to be ---without exception--- of the LINER class, and most
probably {\it not\/} powered by star formation activity. 
I discuss two independent pieces of evidence which suggest
that the extended gas in giant ellipticals is ionized by means of a
distributed source of ionization:\ 
{\it (i)\/} A significant correlation exists between the \Ha+\NII\
luminosity and  the optical luminosity within the region occupied by
the ionized gas, and 
{\it (ii)\/} the ionization parameter of the gas does not change
significantly with galactocentric radius. 
Two potential sources of ionization are evaluated:\ Photoionization by
old hot stars (of post-AGB and/or AGB-Manqu\'e type) and mechanical
energy flux from electron conduction in hot, X-ray-emitting gas. 
\end{abstract}
 
 
\section{Introduction} 

Elliptical galaxies have long been considered to be inert stellar
systems, essentially devoid of interstellar matter. However, 
our understanding of the nature of the multi-phase interstellar medium
(ISM) in elliptical galaxies has undergone a radical change from this 
consensus that prevailed until the mid-1980's. Present-day 
instrumental sensitivity across the electro-magnetic spectrum has
revealed the presence of a complex, diverse ISM in elliptical
galaxies, challenging their very definition in Hubble's galaxy
classification scheme. 

Early-type galaxies are important in regard to any theory on the
evolution of gas in galaxies simply {\it because\/} they generally
don't show any evidence for the presence of any significant amount of cool
interstellar matter, compared to spiral (or irregular) galaxies. In
fact, as of today, many basic physical and evolutionary relationships
between the various components of the ISM in ellipticals are not yet
understood, contrary to the situation in spiral galaxies (for the
latter:\ see \eg Tinsley 1980; Knapp 1990; Brinks 1990; 
Ostriker 1990). A currently particularly
controversial issue in this field of work is the question whether
galaxy interactions or cooling flows dictate the interplay between the
different components of the ISM in giant\footnote{This review
primarily concerns {\sl Giant\/} ellipticals, defined here as
ellipticals with $M_B < -19$ (cf.\ Mihalas \& Binney 1981).}
ellipticals (cf.\ Sparks, Macchetto \& Golombek 1989; de Jong \etal
1990; %Sparks 1992; 
Fabian, Canizares \& B\"ohringer 1994; Goudfrooij
\& Trinchieri 1998). The ``warm'' ($T \sim 10^4$ K) gas component
(most often referred to as  ``ionized gas'' or ``optical nebulosity'') is 
crucial in the context of this controversy because its 
luminosity can dominate ---at least locally---
the total energy output of all (radiating) components of the ISM (see,
\eg Sparks \etal 1989; Goudfrooij 1994). It is therefore important to
understand the nature of the optical nebulosity (\eg is it physically
associated with another component of the ISM\,?; does it trace
star formation activity or is it ionized by other means\,?), in
order to arrive at a correct description of the physics of the ISM in
giant ellipticals. 

The plan of this paper is as follows. I briefly review the 
multi-phase ISM in elliptical galaxies in Section 2. In Section 3, I
discuss the physical relations of the warm gas with other
components of the ISM in ellipticals, along with possible implications
of the observed correlations. Section 4 discusses possible ionization
mechanisms of the warm gas in the light of new results on
emission-line intensity ratios from a low-resolution
spectrophotometric survey of giant ellipticals. 

\section{The Multi-phase ISM of Elliptical Galaxies}

\subsection{What is the Fate of Stellar Mass Loss in Ellipticals\,?} 

The most obvious source of interstellar matter in early-type galaxies
is the material injected into interstellar space by stellar winds
of evolving stars. A good estimate on the steady-state mass injection
rate for an old stellar population typical for elliptical
galaxies is provided by the product of the birth rate of planetary
nebulae with the mass difference between the main sequence turn-off
mass of such old stellar populations ($M_{\scsl{MSTO}}\sim 1.0$ \Mzon) and
the mean mass of white dwarfs; this estimate results in $\dot{M}
\simeq 1.5\; 10^{-11} \; 
(L/{\mbox{L}_{\odot}}) \;  \mbox{M}_{\odot} \; \mbox{yr}^{-1}$, and
should be accurate to within a factor of three (cf.\ Faber \&
Gallagher 1976). Since this mass loss mechanism must have been active
since the formation epoch of giant ellipticals, it has
yielded at least 
$10^{10}\;(L/10^{11}\Lsun)$ \Mzon\ at the present time, assuming a
galaxy age of $\sim$\,$10^{10}$~yr. Note that this actually represents a
strict {\it lower limit} to the accumulated material from stellar mass
loss, as the mass loss rate was much higher during the era of galaxy
formation when the star formation rate was very high. What is the fate
of this material in ellipticals\,? \\ [-5ex]

\paragraph{Cool Gas\,? No\,!} 
With the advent of increasingly sensitive radio-- and mm-wave detector
technologies, several groups searched for this interstellar material
in giant ellipticals through surveys of the usual tracers of cool gas
%which dominates the ISM (in mass) in spiral galaxies 
(\eg Knapp \etal 1985 (\HI); Lees \etal 1991 (CO); Bregman, Hogg \&
Roberts 1992 (\HI\ + CO); Wilkind, Combes \& Henkel 1995 (CO)).
%; Braine, Henkel \& Wiklind 1997 (\HI). 
While \HI\ and CO detections
of ellipticals are becoming more common (see contributions of Knapp
and Oosterloo, this volume), it has 
become clear that these cold gas components are
{\it not\/} tracing the mass lost by 
stars within these galaxies:\ The amount of cold gas observed is orders of
magnitude less than expected from accumulative mass loss, while 
histograms of the ``normalized cold gas content''
($M_{\scrm{H\,}\scsc{i}}/L_B$ and $M_{\scrm{H$_2$}}/L_B$) for ellipticals
show a wide distribution without any clear peak, \ie there seems to be no
causal connection between the cold gas and the stellar content in giant
ellipticals %(cf.\ Knapp \etal 1985; Lees \etal  1991; Bregman \etal 1992). 
(\eg Bregman \etal 1992). 
%
%The same situation
%is encountered for the distribution function of the normalized amount
%of dust ($M_{\scrm{dust}}/L_B$) in ellipticals as derived from {\sl
%IRAS\/} measurements (Knapp \etal 1989; Forbes 1991; Goudfrooij \& de
%Jong 1995). 
%
This is in strong contrast to the situation among spiral
galaxies, where the amount of cold gas is proportional to the galaxy
luminosity with very little scatter in the relationship (\eg Haynes \&
Giovanelli 1984). These findings strongly suggest that the 
cold gas in ellipticals is generally of {\it external\/} origin (\ie
captured during galaxy interactions), and that the gas lost by stars
had been blown out of these galaxies by a hot, supernova-driven wind
as first proposed by Mathews \& Baker (1971).  \\ [-5ex]
%; see also White \& Chevalier 1983). 

\paragraph{Hot Gas\,!}
So what {\it is\/} the fate of the material produced by stellar
mass loss in giant ellipticals\,? This issue finally got substantially
clarified after the first X-ray observations of giant ellipticals with
the {\it Einstein\/} satellite were analyzed. 
Many giant ellipticals were found to be embedded in extended halos of
X-ray emission, and the X-ray colors showed that the emission was
dominated by thermal Bremsstrahlung from hot ($T\sim 10^7$ K) gas, with 
total gas masses of order $10^9 - 10^{11}$ \Mzon\footnote{H$_0$ = 50
\kms\ \Mpc\ is assumed throughout this paper}, quite similar to the expected
amount from accumulated mass loss (Forman, Jones \& Tucker
1985). Evidence for the existence of this hot component of the ISM is
currently only substantial for the most luminous ellipticals ($L_B \ga
5 \; 10^{10}$ \Lzon, see \eg Kim, Fabbiano \& Trinchieri 1992). Thus,
only the most massive ellipticals seem to be able to retain the
material lost by stellar winds and suppress the supernova-driven wind 
proposed by Mathews \& Baker (1971) (see also Loewenstein, this
volume). In this scenario, smaller ellipticals with 
too shallow potential wells may not see the bulk of their internally
produced ISM ever again, donating it to the intracluster (or
intergroup) medium.  

\subsection{So what about the ``Warm'' Ionized Gas Component\,?}

\subsubsection{2.2.1. Optical Surveys} \ \\ [-2.5mm]

\noindent
``Warm'' ionized gas was the first component of the ISM known to exist
in a number of apparently ``normal'' ellipticals. The first notion of
the frequency with which optical emission lines occur in ellipticals
was provided by Mayall (1958) who examined the redshift catalog of
Humason, Mayall \& Sandage (1956) and found an \OII\,\lda3727 
detection rate of 12\%\ for E galaxies (and 48\%\ for S0
galaxies). More recent long-slit spectroscopic surveys of
early-type galaxies (Caldwell 1984; Phillips \etal 1986) pushed the
detection rate up to about 50--55\%. Phillips \etal observed the
\Ha\,\lda6563 \&\ \NII\,\lda\lda6548,\,6583 emission lines for a large
sample of 203 E and S0 galaxies and found that the emission-line
spectra of ellipticals come in two distinct flavors:\ {\it (i\/)}
narrow line widths (FWHM $\la$ 200 \kms) and a small ($\la 0.4$)
\NII\,\lda6583/\Ha\ intensity ratio, reminiscent of normal
galactic \HII\ regions; these characteristics are found in {\it
low-luminosity\/} ellipticals featuring blue galaxy colors, and {\it
(ii)\/} relatively broad lines (FWHM $>$ 200 \kms) and a large ($\ga
1.0$) \NII\,\lda6583/\Ha\ ratio, which is typical for the 
so-called LINER class (Low-Ionization Nuclear Emission-line Regions,
Heckman 1980); the latter characteristics are typically found in {\it
giant\/} ellipticals.  

While giving important information on the physical nature of the
ionized gas, long-slit spectroscopy does not allow one to appreciate
the full spatial extent of the emission-line gas. 
%so that \eg derived emission-line luminosities from such spectra are
%not very useful quantities in a statistical sense. 
With this in mind,
a number of teams have undertaken narrow-band CCD {\it imaging\/}
surveys of elliptical galaxies by isolating the most prominent optical
emission lines, \Ha\ and the \NII\,\lda\lda6548,\,6583 doublet. Kim (1989)
observed 26 ellipticals and S0s detected by {\sl IRAS}, Shields (1991)
observed 46 galaxies detected at X-ray wavelengths, Trinchieri \& di
Serego Alighieri (1991) observed 13 X-ray-emitting
ellipticals that are not at the center of clusters, Buson \etal (1993)
observed 15 ellipticals already known to exhibit emission lines in
their spectra, Goudfrooij \etal (1994) observed a optically complete
sample of 56 ellipticals from the Revised Shapley-Ames (RSA) catalog, Singh
\etal (1995) observed 7  X-ray bright ellipticals and S0s, and
Macchetto \etal (1996) observed a sample of 73 ellipticals and S0s
representing a broad variety of X-ray, radio, far-IR and kinematical
properties. These surveys have shown that the ionized gas in
ellipticals typically has an extended distribution (often out to
several kpc from the center), and can have a 
plethora of different morphologies: smoothly following the stellar
isophotes, flattened disk-like structures, or filamentary structures
that do not resemble the underlying stellar isophotes whatsoever. 

\subsubsection{2.2.2. Relations of the Warm Ionized Gas with other Galaxy
  Properties} \ \\ [-2.5mm]

\noindent 
In the light of the results of the narrow-band imaging surveys
mentioned above, I will discuss relations between the warm ionized gas
component and other relevant (stellar as well as interstellar)
properties of ellipticals in the following paragraphs. 
%Part of these
%relations were discussed before in Goudfrooij (1997), but included
%here for reasons of completeness. 
In case of 
ellipticals that were imaged in more than one of the narrow-band
surveys, I have chosen to use the flux from the deepest survey
available, which typically reported higher fluxes than the other
one(s). This is most likely due to the fact that many ellipticals
contain extended emission-line gas at low surface brightness which the
shallower surveys were not able to detect. Galaxies in which no ionized
gas was detected were assigned upper limits according to the 3-sigma
detection limit to the emission-line surface brightness, assuming a
radial extent of 1 kpc. All luminosities were derived (or converted) 
using $D_N\,$--$\,\sigma$ distances from Faber \etal (1989).  \\ [-5ex]

%\paragraph{Ionized Gas vs.\ Radio Continuum Emission}
%{\sf (Maybe just short referral to Goudfrooij 1997 in next section)}

\begin{figure}[t]
\centerline{
\psfig{figure=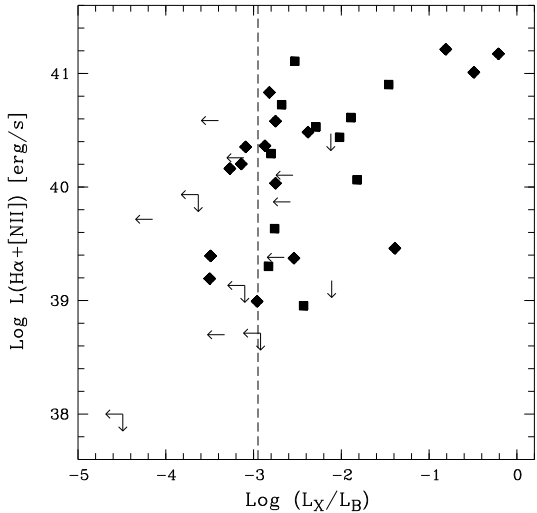,width=7.cm,angle=-90.}
}
\caption[]{\baselineskip=0.9\normalbaselineskip
\Ha+\NII\ luminosity vs.\ X-ray--to--blue luminosity ratio
for ellipticals detected by {\it EINSTEIN}. Filled squares are data
from Macchetto \etal (1996) or Trinchieri \& di Serego (1991); filles
lozenges are data from Goudfrooij \etal (1994). The dashed line
depicts the ration $L_X/L_B$ below which the X-ray emission is
{\it not\/} primarily due to hot gas (Kim \etal 1992)}
\label{f:HaNII_LXLB}
\end{figure}

\paragraph{Ionized Gas vs.\ Hot, X-ray-emitting Gas}
Phillips \etal (1986) were the first to draw the attention to a
possible relation between the warm ionized gas and the hot,
X-ray-emitting gas in ellipticals, as was already known to be the case
for central dominant cluster galaxies featuring ``cooling flows'' (Hu,
Cowie \& Wang 1985; see also Heckman \etal 1989; Fabian, Nulsen \&
Canizares 1991): They found emission-line gas in 12 out of 14
ellipticals which were found by Forman \etal (1985) to contain
significant amounts of hot gas.  
This finding prompted two teams to investigate this possible relation 
further by means of a narrow-line imaging survey of X-ray-emitting
ellipticals. However, they came up with different conclusions: Shields
(1991; using a 1-m telescope) found essentially {\it no\/}
correlation between the \Ha\ and X-ray luminosities, whereas
Trinchieri \&\ di Serego (1991; using a 3.6-m 
telescope) found that galaxies with a relatively large amount of hot
gas also have more powerful line emission {\it on average}, albeit
with quite considerable scatter. This comparison illustrates the
importance of establishing a {\it deep\/} emission-line surface
brightness detection threshold in this line of work. After adding the
results from the deep surveys of Goudfrooij \etal (1994) and Macchetto
\etal (1996), we show in Fig.\ \ref{f:HaNII_LXLB} the relation of the
\Ha+\NII\ luminosity with the X-ray--to--B-band luminosity ratio 
($L_X/L_B$), as it currently stands. 
We use the quantity $L_X/L_B$ rather than $L_X$ to isolate the X-ray
luminosity from hot gas only: The X-ray emission from stellar sources
scales linearly with optical luminosity (as seen in spiral galaxies
without active nuclei, cf.\ Kim \etal 1992), so that $L_X/L_B$ scales
with hot gas content above the typical value observed in spiral
galaxies ($-$2.95 in the log, cf.\ Fig.\ \ref{f:HaNII_LXLB}). 

While Fig.\ \ref{f:HaNII_LXLB} does reveal a positive trend in the
sense that galaxies with larger hot gas content typically 
have high \Ha+\NII\ luminosities as well, the scatter in the
relation is significant:\ some galaxies with high $L_X/L_B$
are weak \Ha\ emitters, and many ellipticals that have bright, extended 
\Ha\ emission have X-ray luminosities that are only barely above the
expected contribution from stellar sources. In other words, the warm
gas/hot gas connection is clearly more complex than the
``standard cooling flow theory'' (\eg Fabian \etal 1991) would
predict. The depth of the potential well may be reflected by the amount
of hot gas retained in a reservoir of hot gas surrounding the
elliptical (cf.\ Section 2.1), but to contain relatively large amounts
of warm ionized gas at the present time, it does {\it not\/} seem required to
contain such a deep potential well. 
Note that active nuclei are {\it not\/} powering the
ellipticals with high \Ha\ luminosity and low $L_X/L_B$:\ their radio
power is typically orders of magnitude below those of typical radio galaxies
(even those of class Faranoff-Riley I), and their \Ha\ flux does not
correlate with radio (5 GHz) flux (Goudfrooij 1997). \\ [-5ex]

\begin{figure}[t]
\centerline{\psfig{figure=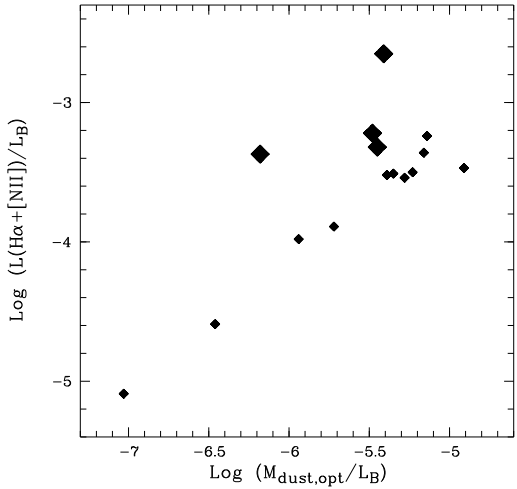,width=7.cm,angle=-90.}}
\caption[]{\baselineskip=0.9\normalbaselineskip
\Ha+\NII--to--B-band luminosity ratio vs.\ ratio of dust
mass and B-band luminosity of dusty ellipticals in the survey of
Goudfrooij \etal (1994). Large symbols are ellipticals in which the
\Ha+\NII\ emission has a large radial extent, out to beyond the radius where
dust absorption could be detected. Note the clear correlation.}
\label{f:HANIILB_MDOLB}
\end{figure}
\begin{figure}[th]
\centerline{
\psfig{figure=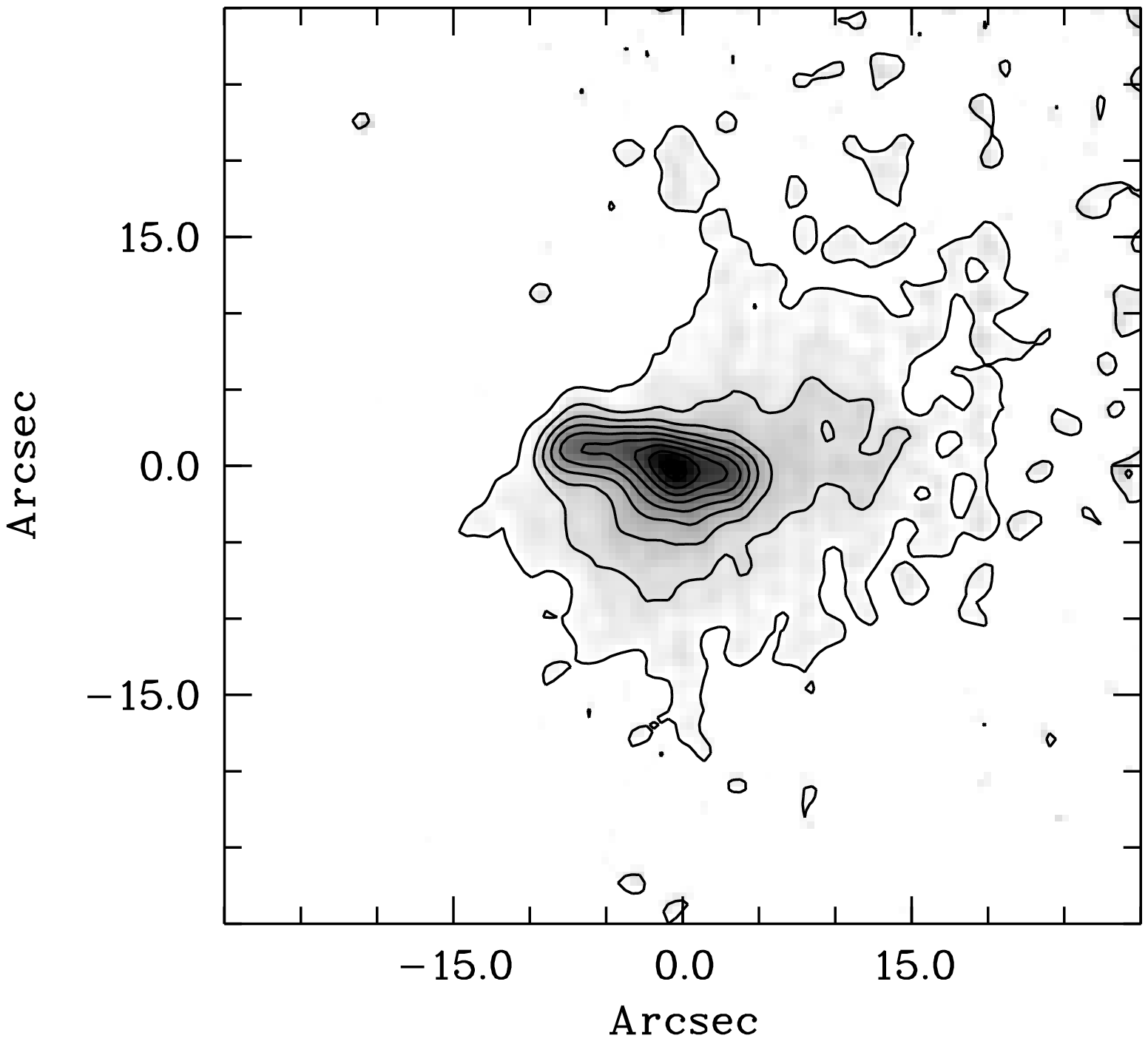,bbllx=55pt,bblly=93pt,bburx=505pt,bbury=505pt,width=6.35cm}
\hfill 
\psfig{figure=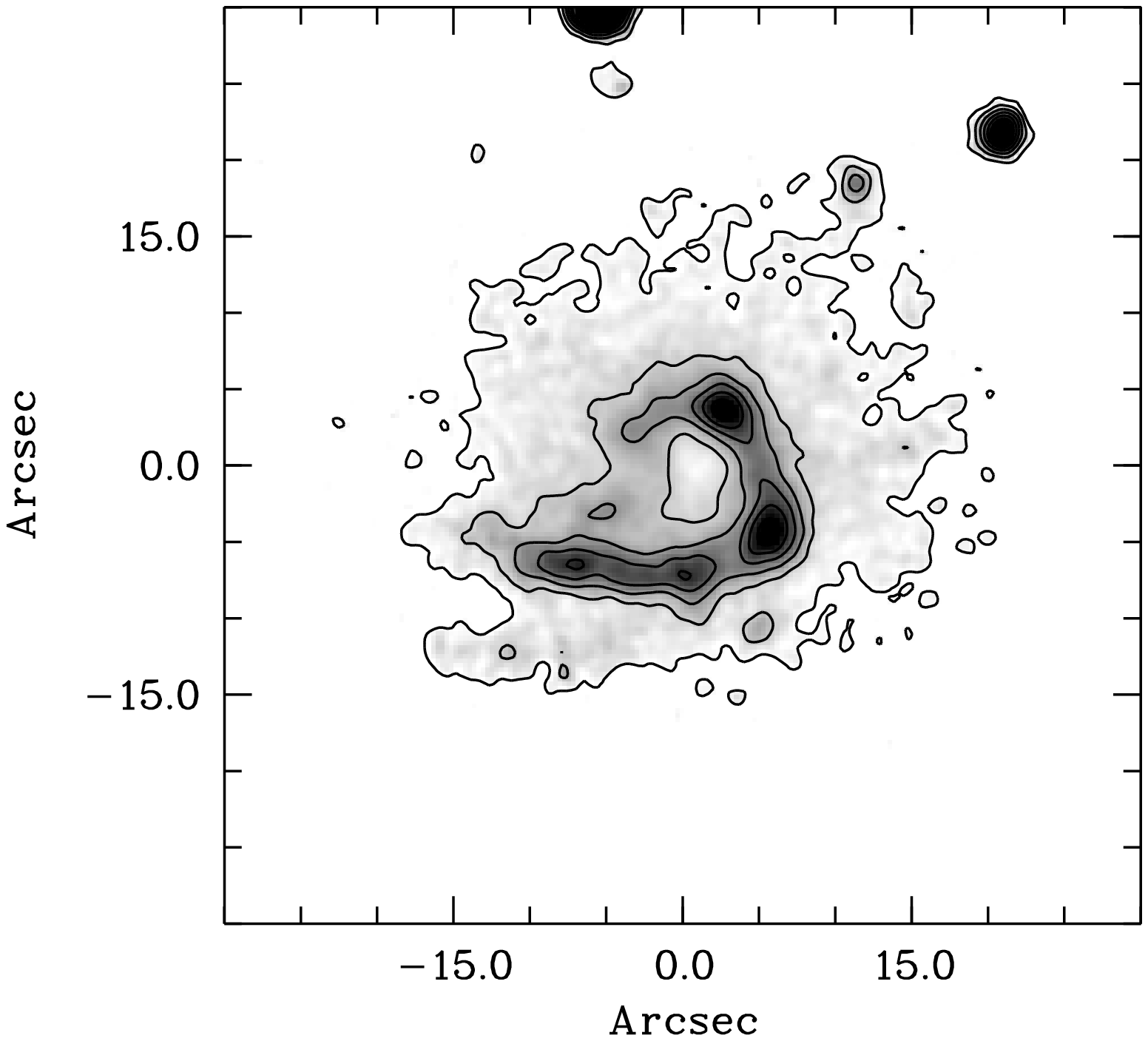,bbllx=55pt,bblly=93pt,bburx=505pt,bbury=505pt,width=6.35cm}
}
\vspace*{-58.13mm}
\centerline{
\psfig{figure=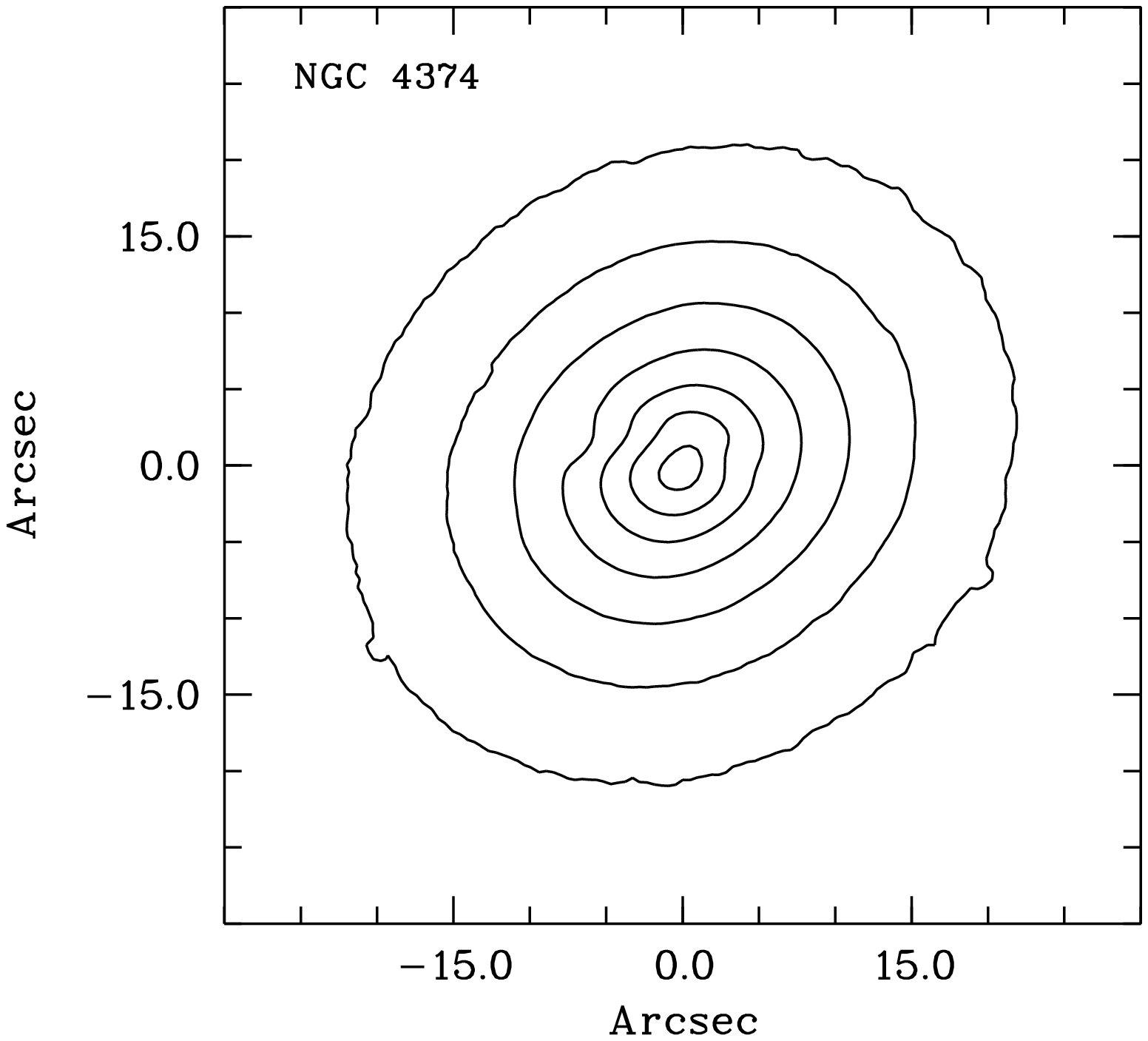,bbllx=55pt,bblly=93pt,bburx=505pt,bbury=505pt,width=6.35cm}
\hfill
\psfig{figure=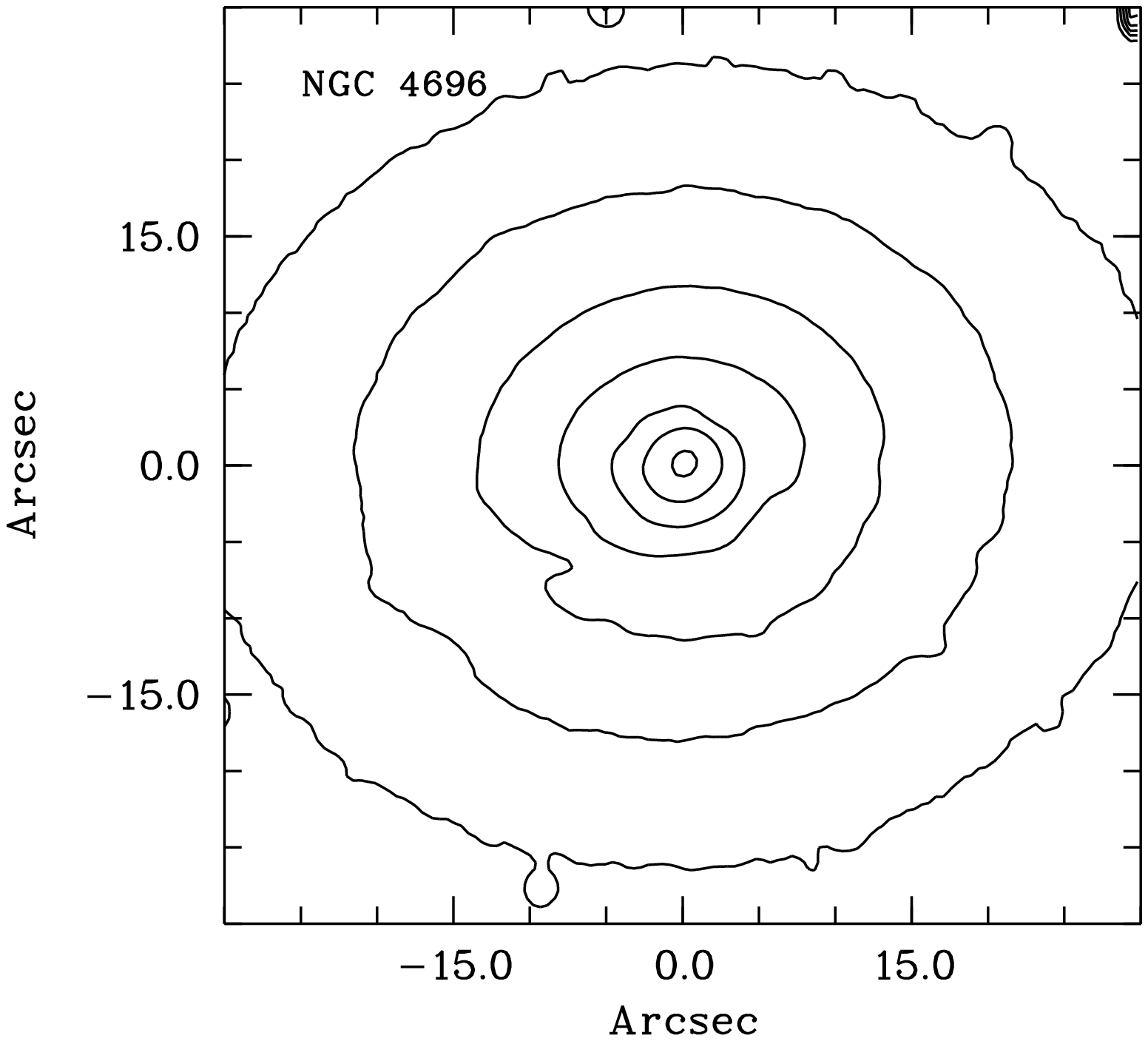,bbllx=55pt,bblly=93pt,bburx=505pt,bbury=505pt,width=6.35cm}
}
\vspace*{0mm}
\centerline{
\psfig{figure=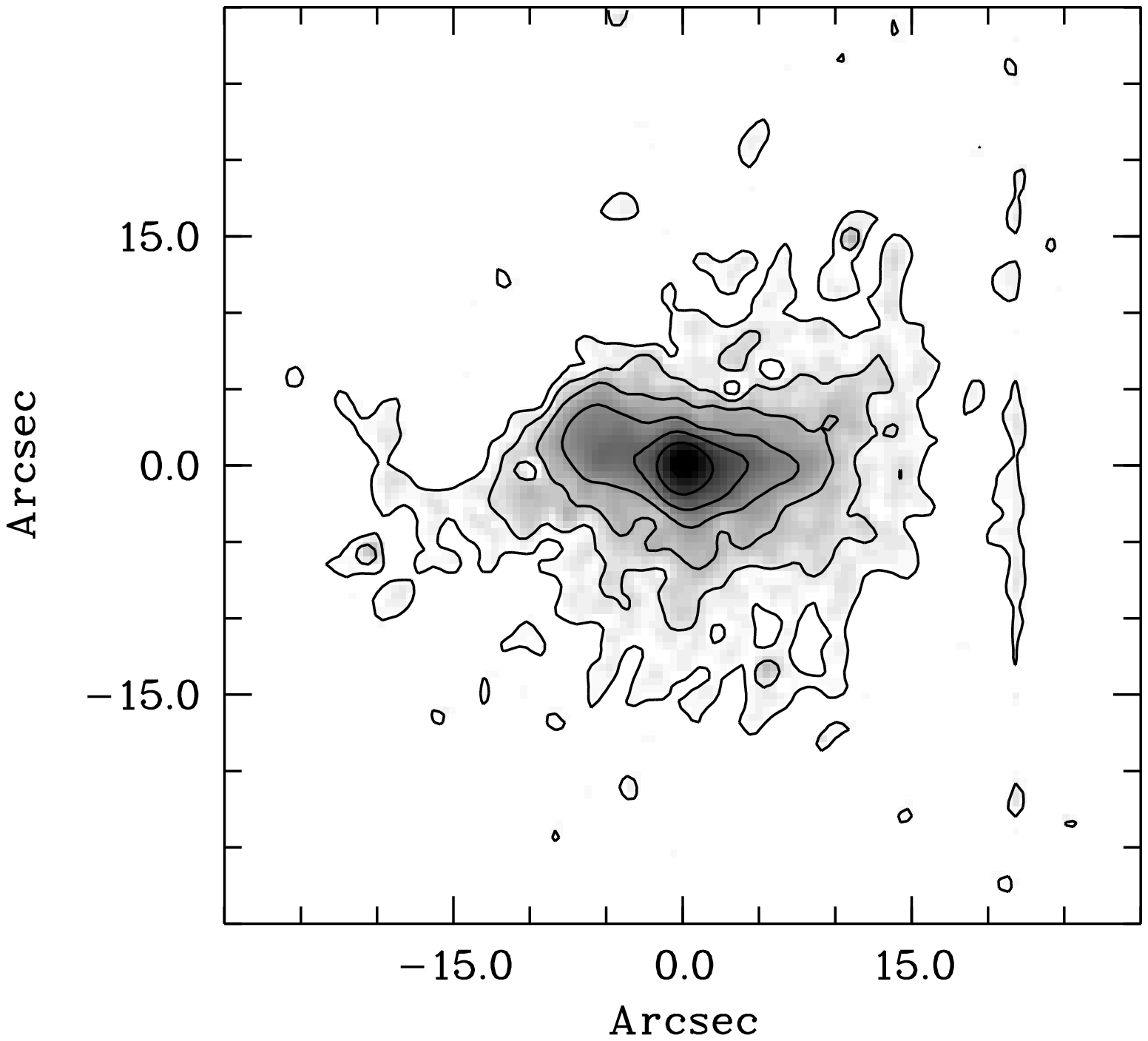,bbllx=55pt,bblly=93pt,bburx=505pt,bbury=505pt,width=6.35cm}
\hfill
\psfig{figure=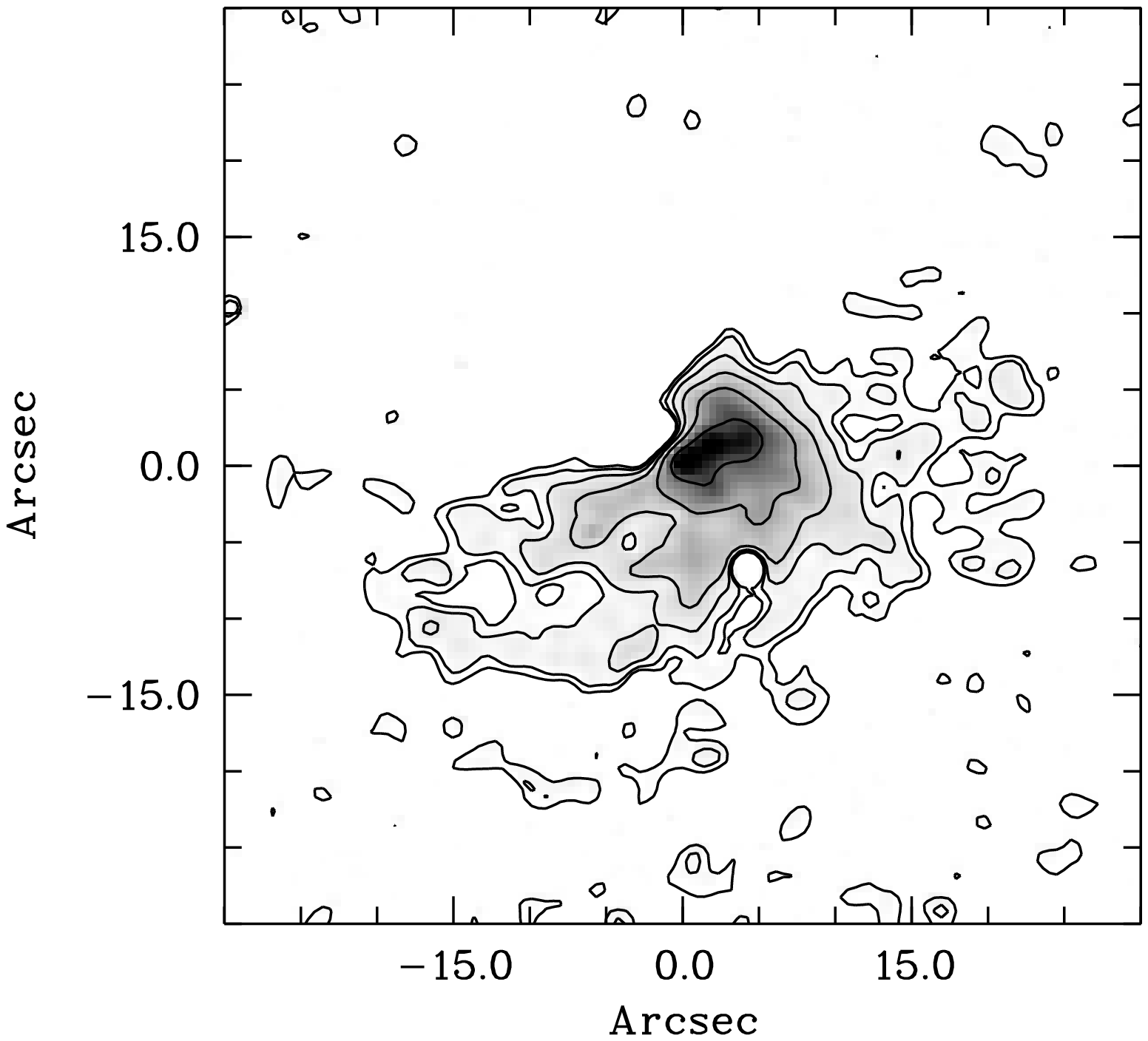,bbllx=55pt,bblly=93pt,bburx=505pt,bbury=505pt,width=6.35cm}
}
\vspace*{-2mm}
\caption[]{\baselineskip=0.9\normalbaselineskip
CCD images of X-ray bright elliptical galaxies. The top
panel shows a $B\!-\!I$ color-index image (grey scales) with isophotal
contours of the $B$-band image superposed (thick solid lines). The
bottom panel shows grey-scales of the \Ha+\NII\ emission (with the
stellar continuum subtracted). {\bf Left}: NGC 4374. 
%$B\!-\!I$ grey scales: 2.15 -- 2.40 mag; $B$-band isophotes:
%20.5 -- 17.5 mag; \Ha+\NII\ grey scales: (0.75 -- 48) 10$^{-16}$
%\ergscm\ arcsec$^{-2}$. 
{\bf Right}: NGC 4696. 
%$B\!-\!I$ grey scales: 2.34 -- 2.54 mag; $B$-band isophotes:
%22.5 -- 19.5 mag; \Ha+\NII\ grey scales: (0.45 -- 14.4) 10$^{-16}$
%\ergscm\ arcsec$^{-2}$. 
Images taken from Goudfrooij \etal (1994).
}
\label{f:fig4374_4696}
\end{figure}
\begin{figure}[th]
\centerline{
\psfig{figure=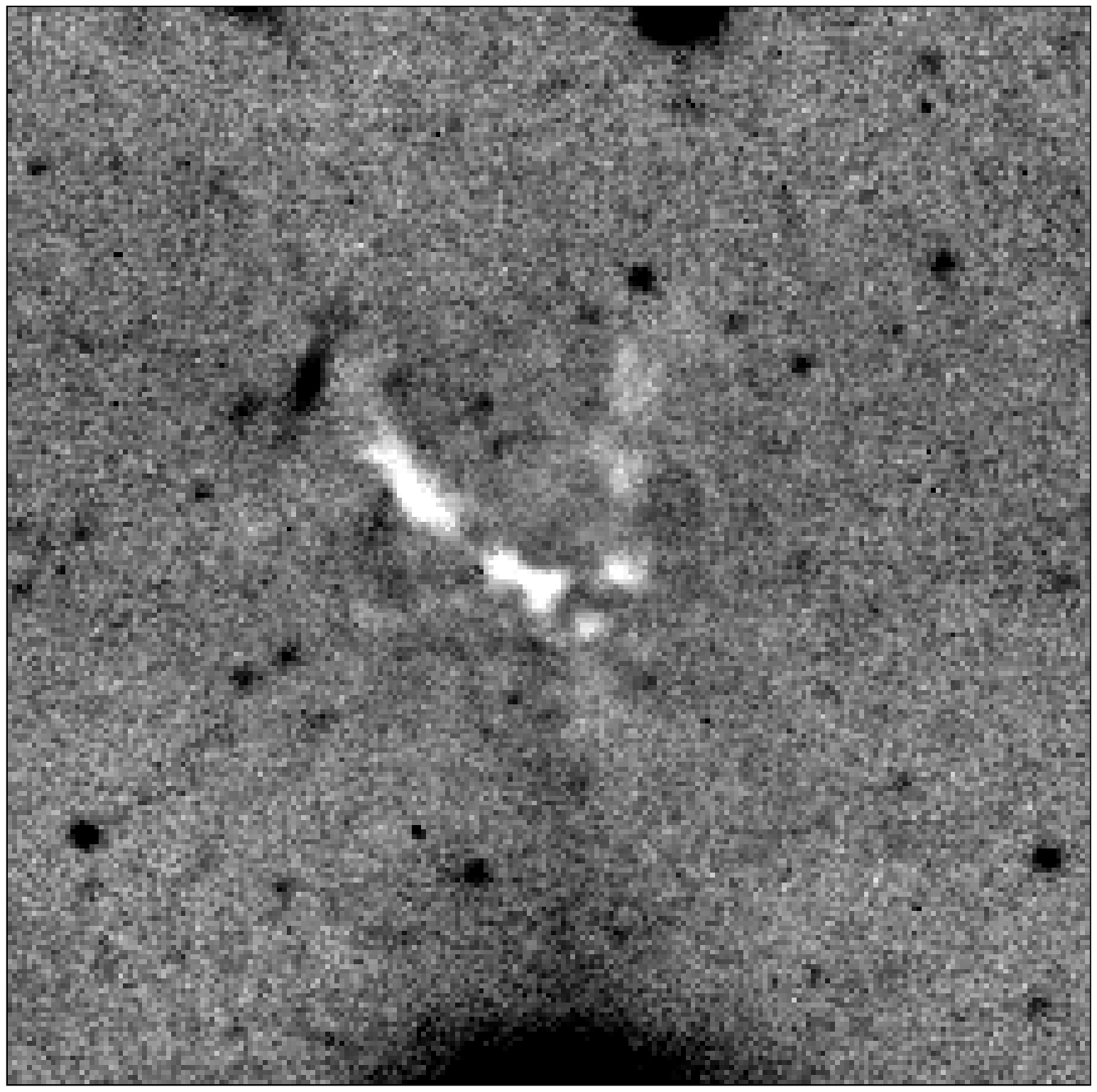,height=6.4cm}
\hspace*{0.05mm}
\psfig{figure=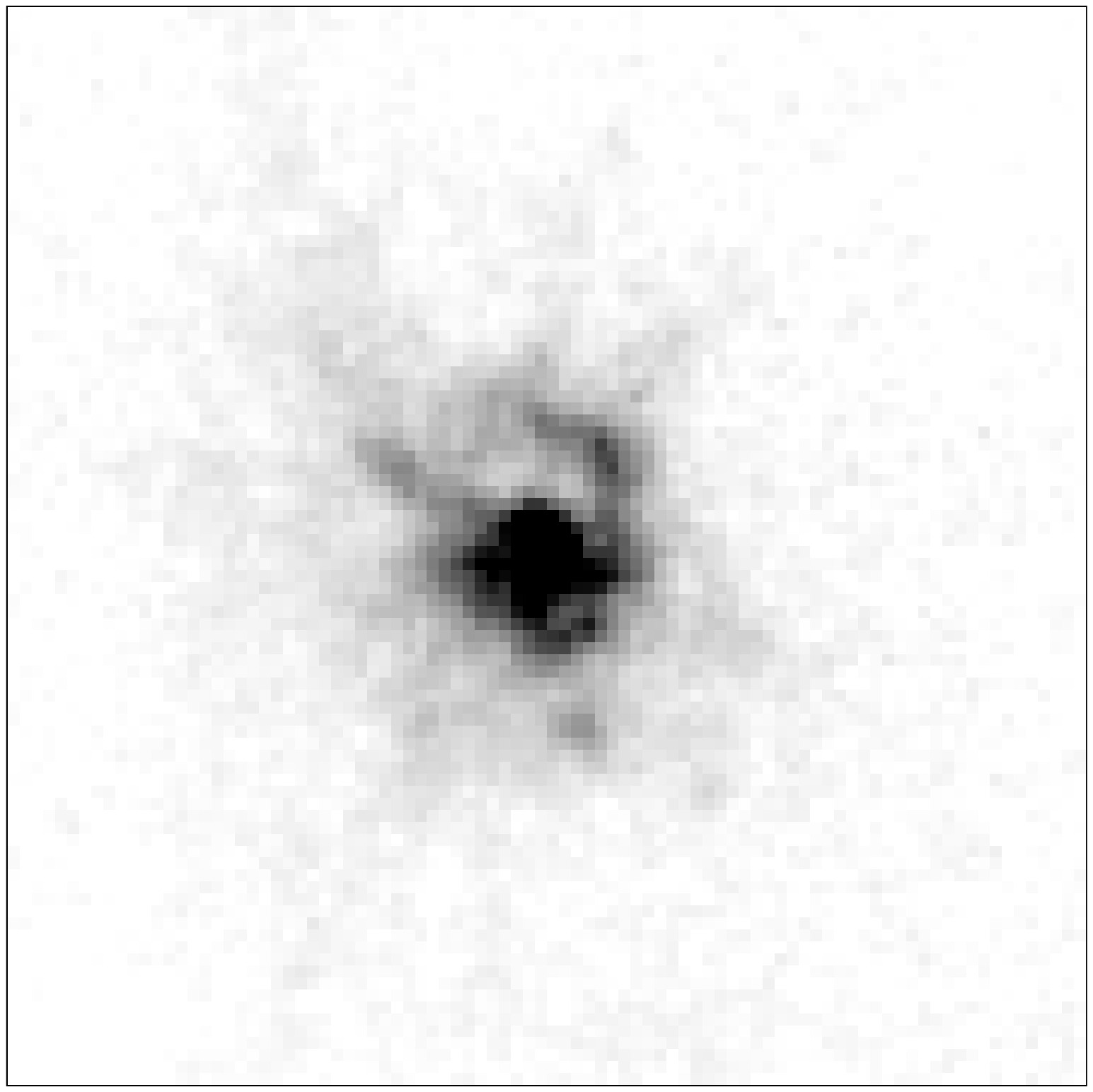,height=6.4cm}
}
\caption[]{\baselineskip=0.9\normalbaselineskip
{\bf Left:} Grey-scale reproduction of the distribution of
$A_V$ of dust extinction in the central $60\times60$ arcsec of
NGC~5846. Maximum $A_V$ = 0.065 (white); detection limit for dust
features is $A_V \sim 0.015$. {\bf Right:} Grey-scale reproduction
of the distribution of \Ha+\NII\ emission in the central $60\times60$
arcsec of NGC~5846. Figure taken from Goudfrooij \&
Trinchieri (1998). 
}
\label{f:fig5846}
\end{figure}

\paragraph{Ionized Gas vs.\ Dust}
The imaging surveys of Kim (1989) and Goudfrooij \etal (1994) revealed
that warm ionized gas in ellipticals is virtually always 
morphologically associated with dust absorption. For instance,
Goudfrooij \etal found \Ha\ emission in 21/22 dusty ellipticals
and dust in 21/31 \Ha--emitting ellipticals, which is
consistent with a one--to--one correlation given the selection effects
(dust is undetectable in (close to) face-on
configurations). This is illustrated by the fact that the mass of dust
in the dust features correlates very well with the \Ha+\NII\
luminosity (see Fig.~\ref{f:HANIILB_MDOLB}). 
Interestingly, this ionized gas/dust connection seems to be
independent of the hot gas content of ellipticals, \ie dust features
are found in X-ray bright ellipticals just as well (see
Figs.~\ref{f:fig4374_4696}--\ref{f:fig5846}). 
This finding provides an important constraint for the origin of the
warm ionized gas in ellipticals because {\it if it would originate
within a ``cooling flow'', it should generally not be associated
with dust}. The lifetime of a dust grain of radius $a$ against
collisions with hot protons and $\alpha$-particles (``sputtering'') in
a hot gas with $T_{\scrm{e}} \sim 10^7$ K is $\tau_{\scrm{d}} =
2\;10^5\; (n_{\tinm{H}}/\mbox{cm}^{-3}) \, (a/0.1\,\mu\mbox{m})$ yr
(Draine \&\ Salpeter 1979; Tielens \etal 1994), 
which is typically of order only $\sim\,$10$^7$ yr for grains of
radius 0.1 $\mu$m (and proportionally shorter for smaller grains) in
the central few kpc of X-ray bright ellipticals where the typical
density is in the range 0.03 -- 0.1 cm$^{-3}$ (see, \eg Trinchieri \etal
1997). Hence, any matter that condenses out of a cooling flow in
the central regions of early-type galaxies is very likely to be devoid
of dust. 
The cooling flow gas is also unlikely to generate
dust internally: While pressures in the central regions of a cooling
flow ($nT \sim 10^5 - 10^6$ cm$^{-3}$\,K) are high compared
with typical pressures in the diffuse ISM in our Galaxy, they are
still significantly lower than those of known sites of grain formation
such as the atmospheres of red giant stars ($nT \sim 10^{11}$
cm$^{-3}$\,K; Tielens 1990). 

To explain the association of dust with warm ionized gas in X-ray
bright ellipticals, Sparks \etal (1989) and de Jong \etal (1990) {\it
independently\/} proposed an alternative to cooling flows:\
the so-called ``evaporation flow'' scenario, in which the
dusty filaments arise from the interaction of a small gas-rich galaxy
with the giant elliptical, or from a tidal accretion event in which
the gas and dust is stripped from a passing normal spiral. 
In this scenario, thermal interaction (heat conduction) between the
cool accreted gas and dust and the pre-existing hot gas locally cools
the hot gas (mimicking a cooling flow) while heating 
the cool gas and dust, thus giving rise to optical and far-infrared
emission. This scenario predicts that the 
distribution of the X-ray emission follows that of the ionized
gas closely, and that dust can be associated with the ionized
gas, being gradually replenished by evaporation of cool molecular
clouds brought in during the interaction (see also Goudfrooij \&
Trinchieri 1998). Furthermore, the metallicity of the ionized gas 
would be unrelated to those of the stars and the X-ray-emitting gas in
this scenario.  

The observed dust/ionized gas association can be regarded as strong
evidence in favor of the ``evaporation flow'' scenario; we will be
able to test the predicted X-ray/optical association in the near
future with {\sl AXAF\/} data which will combine arcsec-scale spatial
resolution with adequate energy resolution.  

\paragraph{Ionized Gas vs.\ Stellar Radiation}

\begin{figure}[t]
\centerline{
\psfig{figure=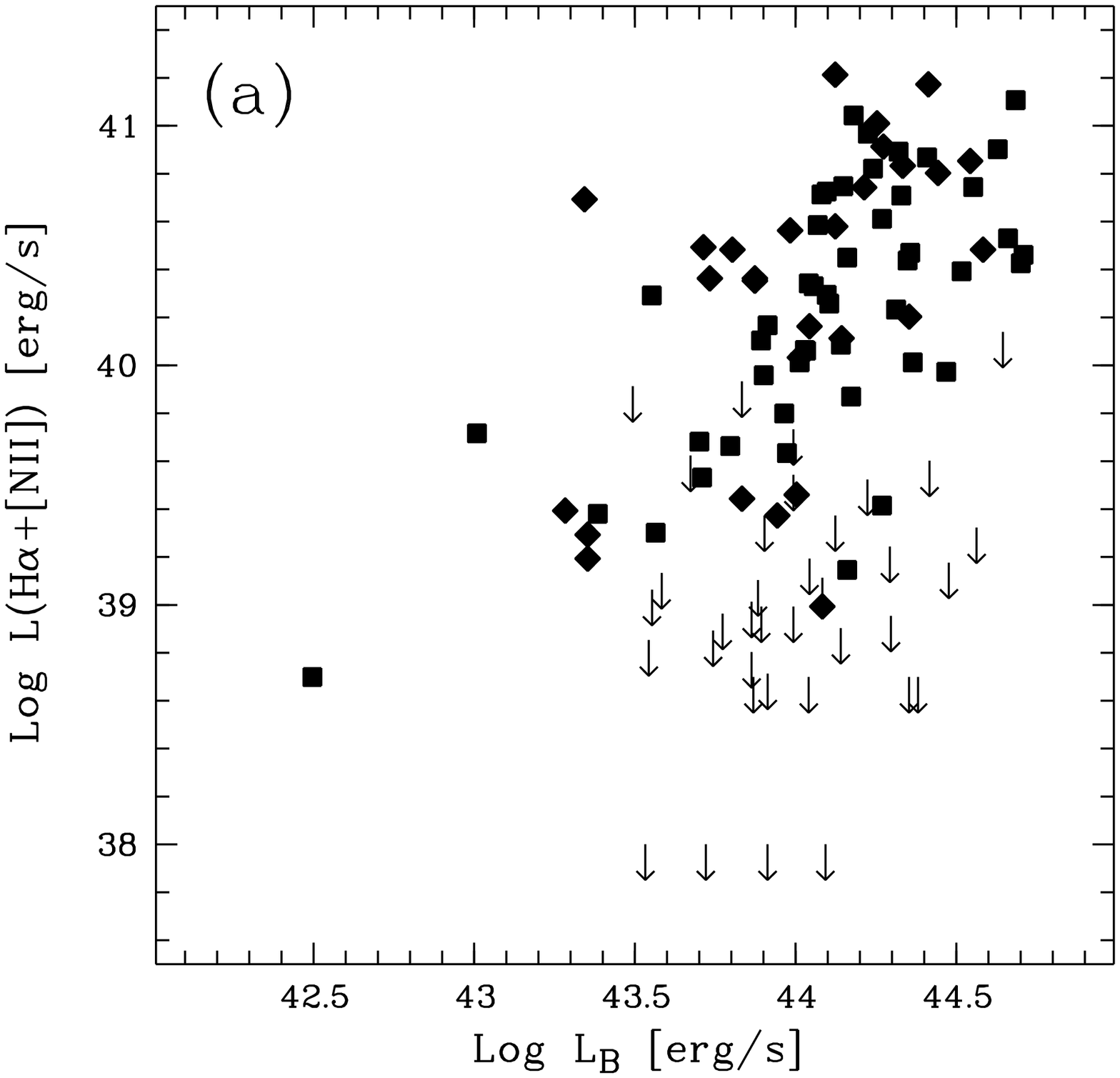,width=6.4cm,angle=-90.}
%\hspace*{0.1mm}
\psfig{figure=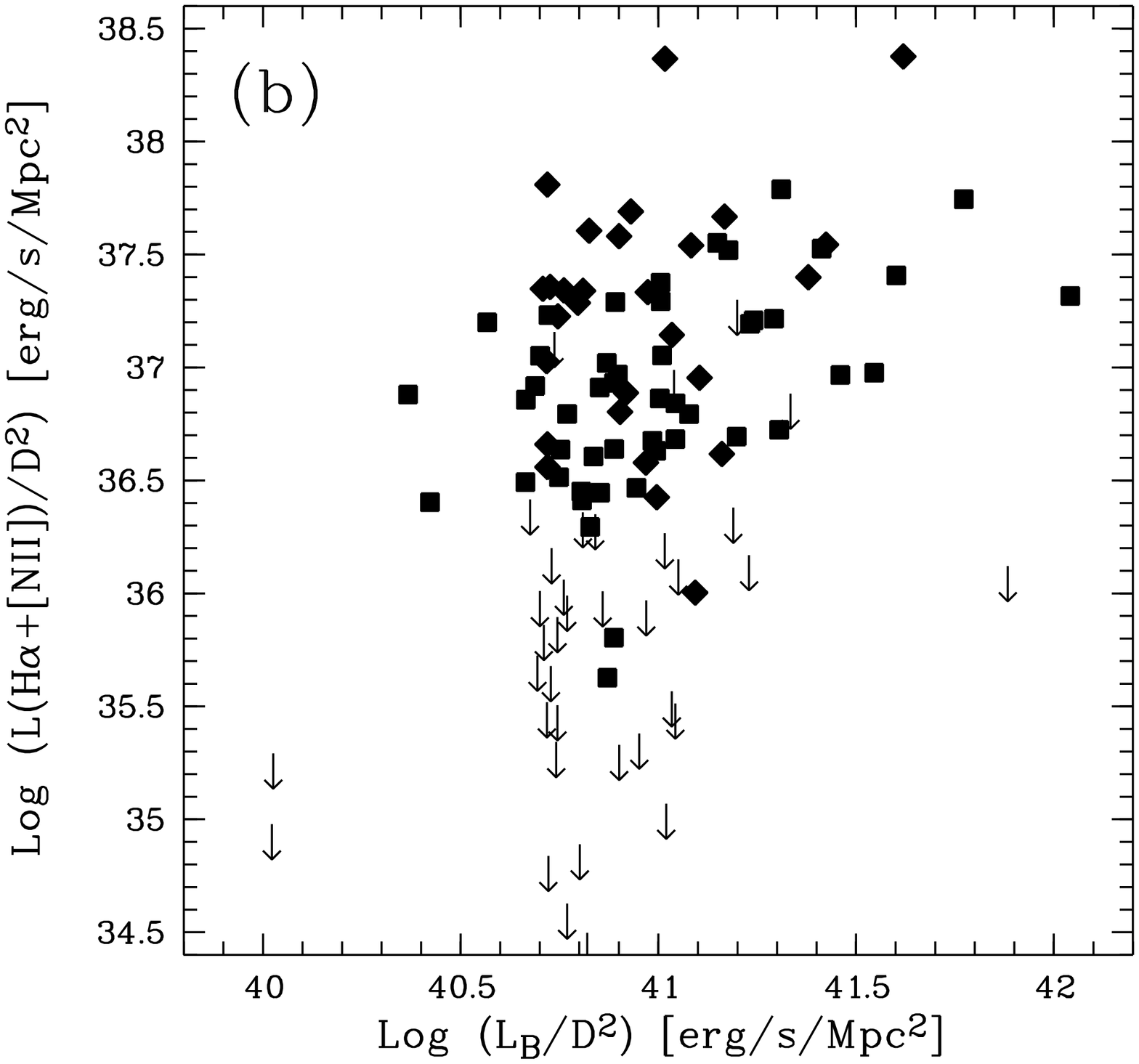,width=6.4cm,angle=-90.}
}
\vspace*{1mm}
\centerline{
\psfig{figure=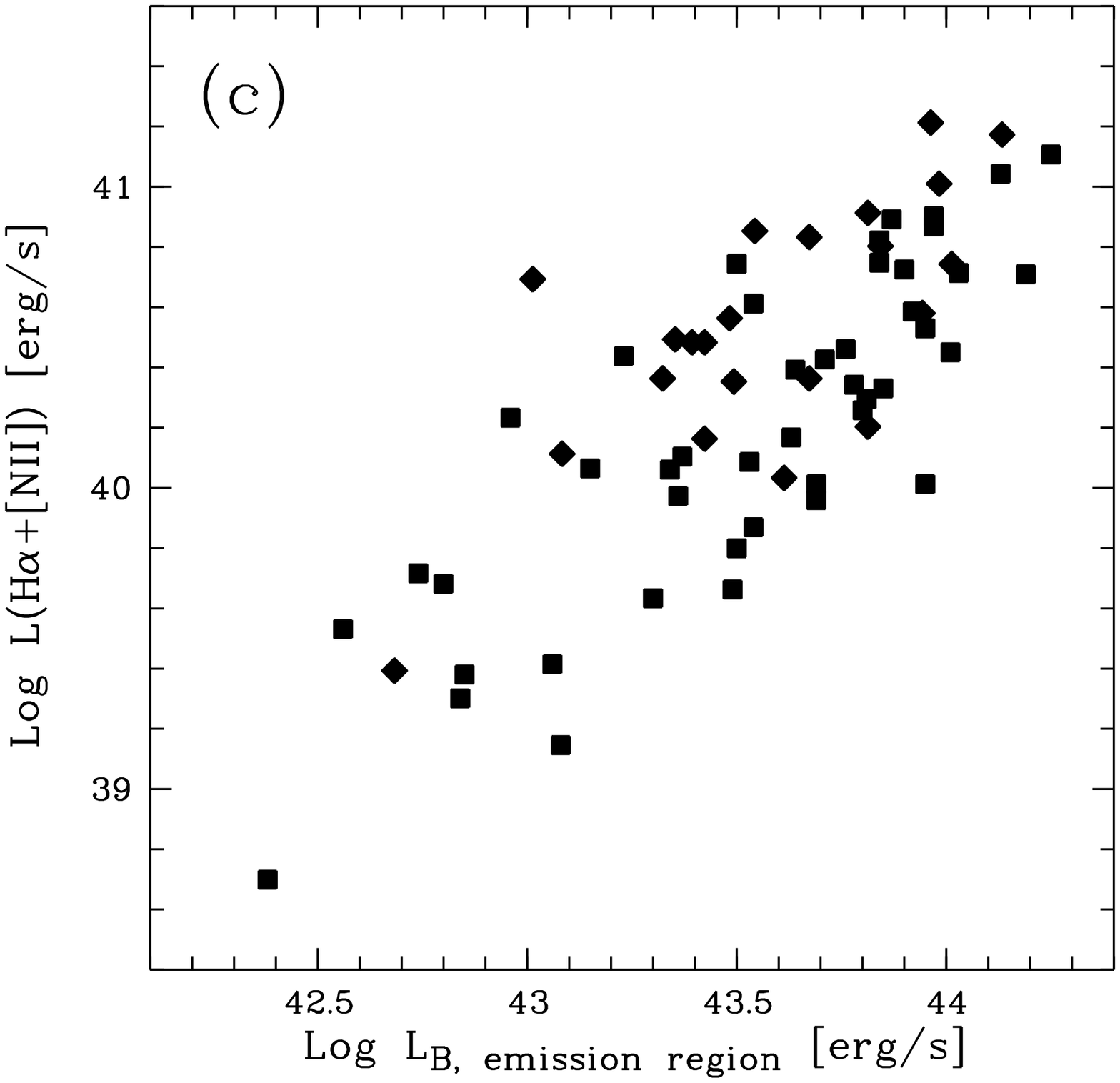,width=6.4cm,angle=-90.}
%\hspace*{0.1mm}
\psfig{figure=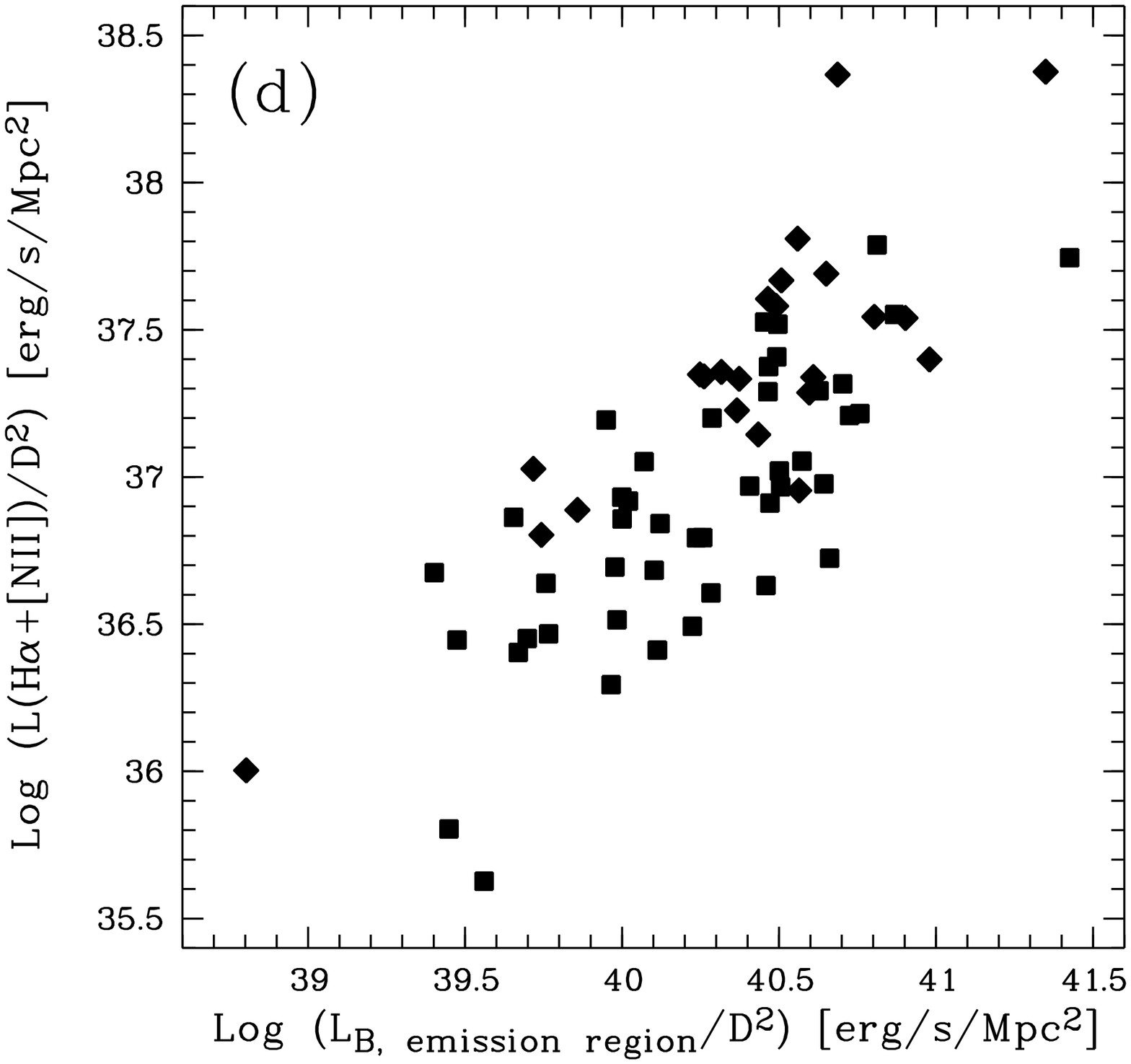,width=6.4cm,angle=-90.}
}
\caption[]{\baselineskip=0.9\normalbaselineskip
Correlations among ellipticals: {\bf (a)} \Ha+\NII\
luminosity vs.\ total B-band luminosity; {\bf (b)} \Ha+\NII\ flux vs.\
total B-band flux; {\bf (c)} \Ha+\NII\ luminosity vs.\ B-band
luminosity within the line-emitting region (see text); {\bf (d)}
\Ha+\NII\ flux vs.\ B-band flux within the line-emitting
region. Symbols as in Fig.~\ref{f:HaNII_LXLB}.  
}
\label{f:HANII_LB}
\end{figure}

While there is a weak trend of the H$\alpha$+\NII\ luminosities with the
total B-band luminosities of the galaxies (cf.\
Fig.~\ref{f:HANII_LB}a), it is clear that many 
luminous ellipticals do not show any sign of ionized gas;
furthermore, the trend is largely due to the {\tt [distance]}$^2$ term
in the luminosities, since the trend disappears largely in a flux-flux
plot (Fig.~\ref{f:HANII_LB}b). Keep in mind, however, that the 
ionized gas is only present in the central parts of these galaxies,
and that its distribution varies greatly from one galaxy to another.  
Interestingly, a plot of the H$\alpha$+\NII\ 
luminosities vs.\ the B-band luminosity emitted within the region
occupied by the line-emitting gas\footnote{defined as a circle 
centered on the nucleus, and with radius equal to
$\sqrt{ab}$, where $a$ and $b$ are the semi-major and semi-minor
axis of the area occupied by the line-emitting gas, respectively.}
reveals a much more evident correlation  (Fig.~\ref{f:HANII_LB}c; see
also Macchetto \etal 1996), which does persist in a 
flux-flux plot (Fig.~\ref{f:HANII_LB}d). 
Taken at face value, this correlation
suggests a stellar origin for the ionizing photons, in line with the
recent result of Binette \etal (1994) who found that post-AGB stars
within the old stellar population of ellipticals provide enough
ionizing radiation to account for the observed H$\alpha$ luminosities
and equivalent widths. Following the prescriptions of Binette {\it et
al.}, predicted H$\alpha$ luminosities have been calculated for the
current collection of galaxies. The result is L$_{{\rm H}\alpha,\,
{\rm obs}}$/L$_{{\rm H}\alpha,\,{\rm pred}} = 1.4 \pm 0.6$, which renders the
stellar origin of ionizing photons quite plausible {\it in
general}. This brings us to the final topic of this paper:\ the
ionization properties of the gas in ellipticals. 

\section{Properties of the Emission-line Spectra of Ellipticals} 

\subsection{A new Spectrophotometric Survey}

In this section I present first results on the emission-line spectra
of giant ellipticals from a new, low-resolution (9--12 \AA\ FWHM)
spectrophotometric survey. The galaxy sample was the one Goudfrooij
\etal (1994) used for their imaging survey (the ``RSA sample'':\ all
galaxies of type E from the RSA catalog with $B_T^0 < 12$ mag). An exhaustive
description of the observations, data reduction and analysis is beyond
the scope of this paper, and will be presented separately 
(MNRAS, in preparation). Here I suffice with a brief summary: The long-slit
spectra have been taken with the B\&C spectrograph on the ESO 1.52-m 
telescope (for the southern gelaxies) and with the IDS
spectrograph on the 2.5-m Isaac Newton Telescope of the Royal
Greenwich Observatory (for the northern galaxies). Care was
taken to cover the spectral range 3600$-$7000 \AA\ in each run, to include
the strongest optical emission lines (from \OII\,\lda3727 to
\SII\,\lda6731). Exposure times were long enough to enable reliable
measurements of 
emission-line intensity ratios not only in the nuclear region, but
also for the extended gas. This is important to study the excitation
mechanism of the extended gas (\eg can it be ionized by a
central AGN-like engine, or must the source of ionizing photons be
extended as well\,?) \\ [-5ex]

\paragraph{The Need for Absorption-line Template Spectra} ~ \\ [0.25mm]
Although now quite commonly detected by means of sensitive
narrow-band imagery, the emission-line component in ellipticals is
typically of low equivalent width (typically only a few \AA\ for the strong
\NII\,\lda6583 line, and even less in the outer regions), and severely
entangled with the underlying stellar population. It is therefore
imperative to construct 
suitable stellar population templates in order to isolate the pure
emission-line component. While different methods have been used in the
past to build such population templates [\eg stellar libraries (Keel
1983); integrated star cluster spectra (Bonatto \etal 1989)], I found
that the absorption-line spectra of giant ellipticals typically cannot be
adequately fitted using spectra of Galactic stars or star clusters. The main
reasons for this problem are {\it (i)\/} the high metallicity of giant
ellipticals, especially their nuclei, and {\it (ii)\/} the non-solar
element ratios in ellipticals (\ie high $\alpha$/Fe ratios, see \eg
Gonz\'alez, this meeting). 
Hence, I chose to build templates from ellipticals that do not show
any evidence (from either imaging or spectra) for either
dust features or emission lines. \\ [-5ex]

\paragraph{Building absorption-line templates} $\!\!\!\!$was performed as
follows. After correcting the flux-calibrated spectra for Galactic
foreground extinction, velocities and velocity
dispersions were measured as a function of radius for each sample
elliptical, out to the outer radii of the ionized gas
distribution. After converting the spectra to zero redshift, I
extracted two spectra for each elliptical, namely (a) covering the central
2$''$, and (b) covering the region from 3$''$ to
the outer radius of the detected \Ha+\NII\ emission (for the gassy
ellipticals, that is; for the gas-free ellipticals, the 
extractions covered a similar range in radius). Line-strength
measurements were then performed on all 
resulting spectra. Care was taken to only consider strong 
absorption-line indices free of any contamination by gaseous emission:\ I
considered Ca\,II\,K (\lda3933), CH (\lda 4300, G-band), Mg$_2$
(\lda5175), and Fe (\lda\lda5270,\,5335). For each spectrum of a gassy
elliptical (\ie each set of measured line strengths), the final
template was built by selecting all spectra of gas-free ellipticals of
which the individual line strengths were the same within 2$\sigma$, where
$\sigma$ is the uncertainty of the (individual) line strength
measurement. After convolving the individual selected spectra with 
Gaussians to match the velocity dispersion of the 
gassy elliptical, the final best-fitting template was created by
averaging the selected spectra, using weights proportional to the
variance in the continuum (at 5500 \AA) of the individual
spectra.   

\begin{figure}[th]
\centerline{
\psfig{figure=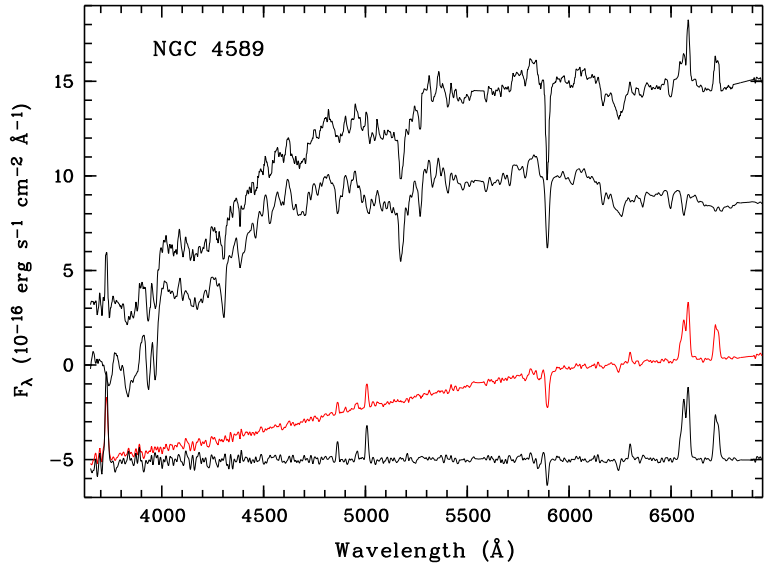,width=10.cm}
\hspace*{1mm}
}
\caption[]{\baselineskip=0.9\normalbaselineskip
Illustration of the absorption-line template subtraction
procedure. The upper spectrum is that of the inner 3$''$ of NGC~4589,
corrected for Galactic foreground extinction and converted to zero
redshift. The second spectrum is
the best-fit template spectrum, normalized to the NGC~4589 spectrum at the
continuum at 4020\,\AA. The third spectrum is the residual, pure
emission spectrum (\ie ``observed''\,$-$\,``template''), which is red due to 
extinction within NGC~4589. The fourth spectrum (bottom) is the
final pure emission spectrum, derived after artifically reddening the
template by \EBV\ = 0.10. The spectra are shifted downwards (relative
to the upper one) by arbitrary amounts along the Y--axis for ease of
visualization. 
}
\label{f:spec4589}
\end{figure}
A typical result of the template fitting procedure is shown in
Fig.~\ref{f:spec4589}. The top spectrum represents the central 3$''$
of NGC~4589, a dust-lane elliptical containing ionized gas. Below that
we show the best-fitting template spectrum. Note that its color is
clearly bluer than that of NGC~4589; this is due to dust absorption
within NGC~4589. The third spectrum shows the result of a direct 
subtraction of the template from the NGC~4589 spectrum, which
illustrates the color mismatch of the two continua once more. We then
correct for the reddening within NGC~4589 by artificially reddening
the template spectrum (using the Galactic extinction curve) until the
continuum slope of the residual spectrum (NGC~4589\,$-$\,template) is
consistent with zero. The final result is shown at the bottom of
Fig.~\ref{f:spec4589}. Note the significant residual
Na\,I\,$D$ absorption at 5890 \AA, which can be attributed to be due to
interstellar matter within NGC~4589. \\ [-5ex]
% (consistent with the presence of a dust lane). 

\begin{figure}[t]
%\centerline{
%\psfig{figure=VO87_OIII_NII.ps,height=6.5cm,angle=-0.25}
%}
%\vspace*{-6.45cm}
%\centerline{\hspace*{7.8mm}
%\psfig{figure=oiiihb_niiha.eps,width=6.8cm,height=5.8cm}
%}
%\vspace*{8mm}
\centerline{
\psfig{figure=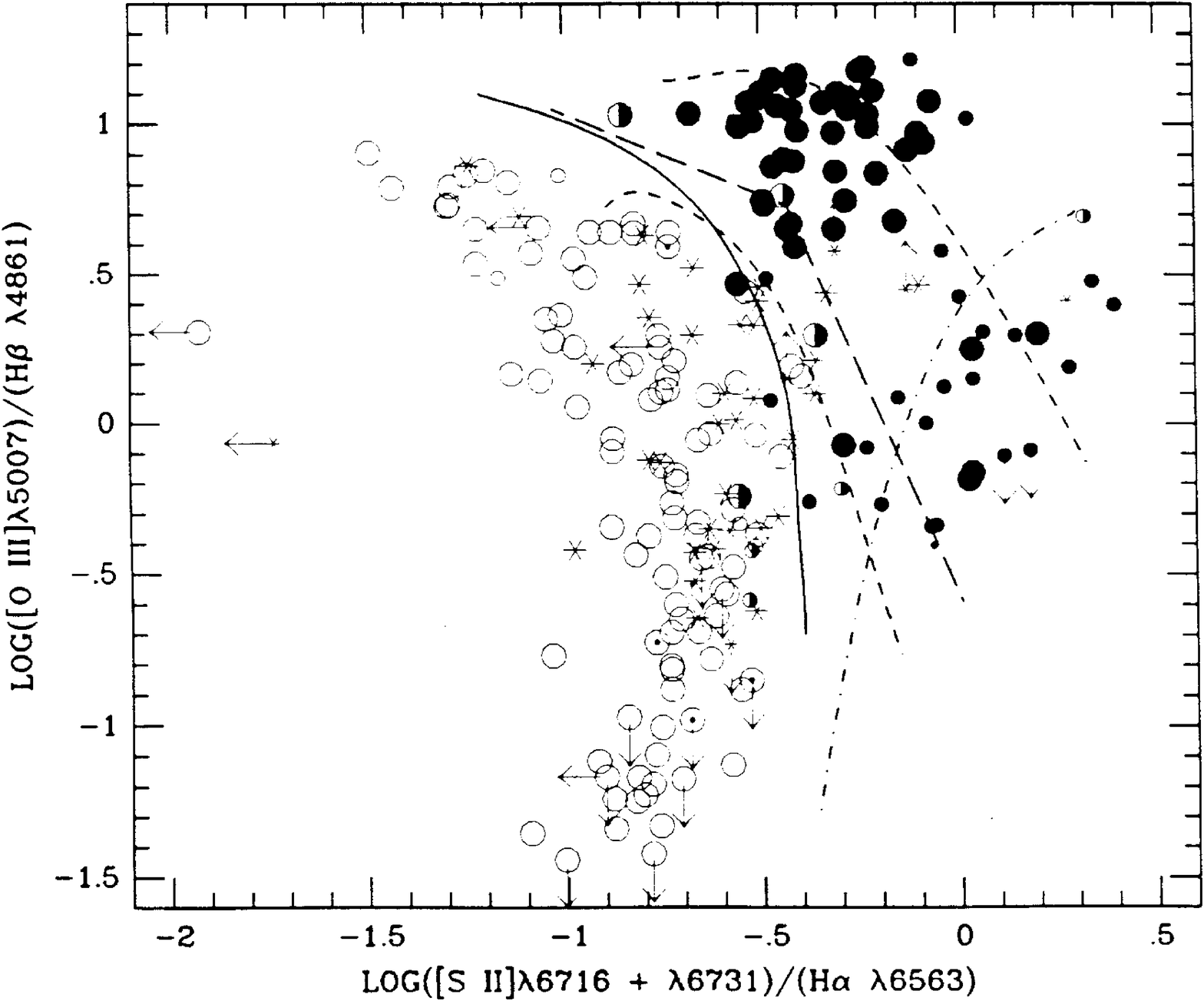,height=5.4cm,angle=-0.3}
\hspace*{1mm} \vspace*{-0.1mm}
\psfig{figure=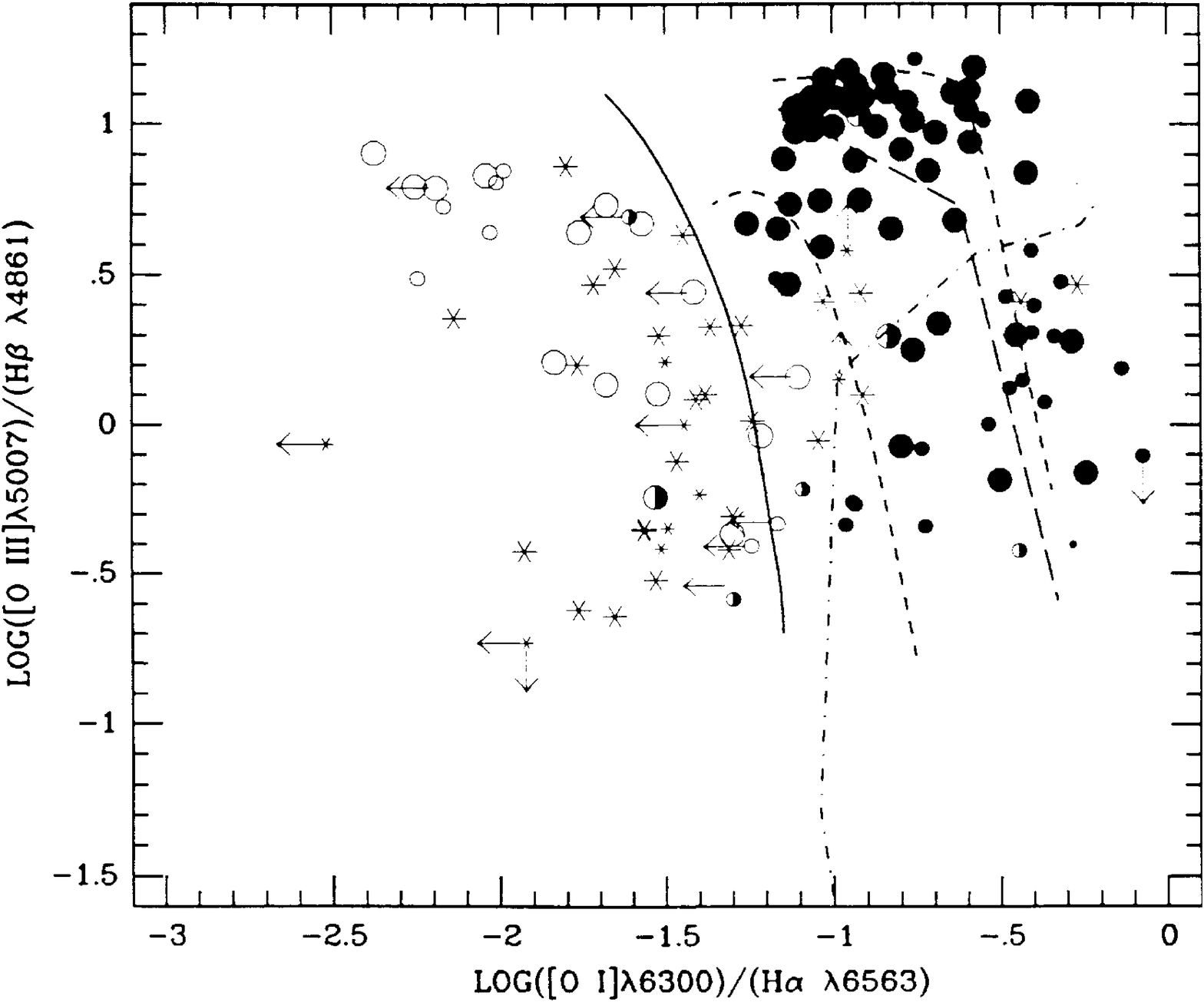,height=5.42cm,angle=-0.1}
}
\vspace*{-5.35cm}
\centerline{\hspace*{7.0mm}
\psfig{figure=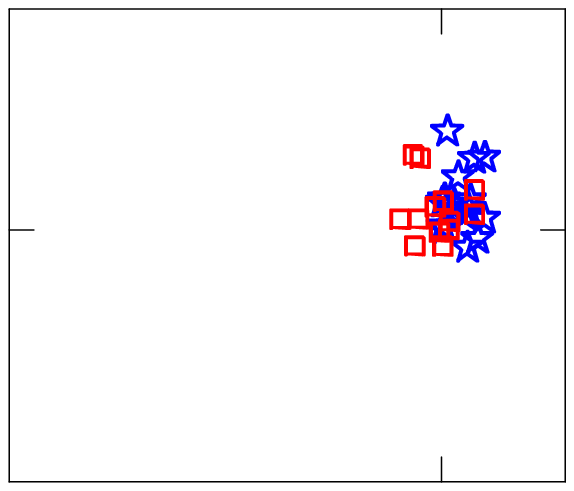,width=5.6cm,height=4.8cm}
\hspace*{9.27mm} \vspace*{0.7mm}
\psfig{figure=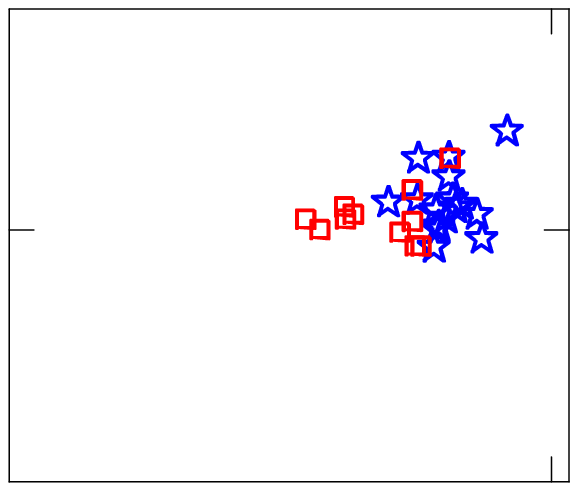,width=5.64cm,height=4.81cm}
}
\vspace*{2mm}
\caption[]{\baselineskip=0.9\normalbaselineskip
Reddening-corrected line ratios for
ellipticals in the ``RSA sample'' (see text), superposed on diagrams
scanned from Veilleux \& Osterbrock (1987). Their symbols
were: {\Large $\bullet$} $\cor$ Seyfert 2, {\scriptsize $\bullet$}
$\cor$ LINER, {\large $\circ$} $\cor$ \HII\ region-like object, and
$\star \cor$ Narrow Emission-Line Galaxy (\ie yet
unclassified). Short-dashed curves represent  
power-law photoionization models from Ferland \& Netzer (1983) for
solar and 0.1 solar abundances (upper and lower,
respectively). Ionization parameter $U$ varies along the curves from
$10^{-1.5}$ to $10^{-4}$. Solid curve divides AGNs from \HII\
region-like objects. 
Open starred pentagons represent data for the 
nuclear 3$''$ of gassy ellipticals in the ``RSA sample''; open
squares represent data for the extended emission-line region of the same
ellipticals. {\rm  {\bf Left:} \OIII\,\lda5007/\Hb\ vs.\
\SII\,(\lda6716\,+\,\lda6731)/\Ha; {\bf Right:}
\OIII\,\lda5007/\Hb\ vs.\ \OI\,\lda6300/\Ha. } 
}
\label{f:VO87plots}
\end{figure}

\begin{figure}[t]
\centerline{
\psfig{figure=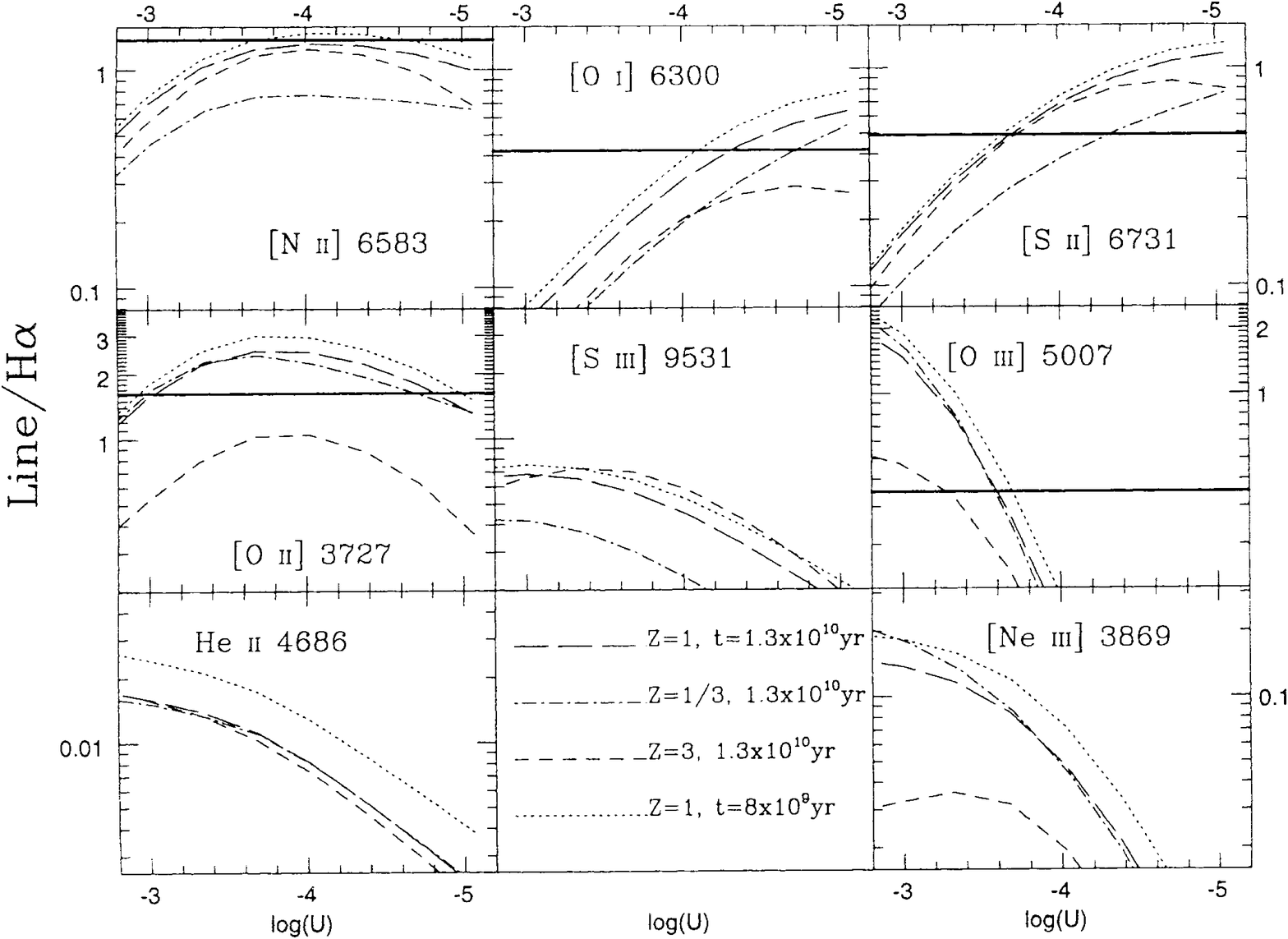,height=8.cm,angle=-0.2}
}
\vspace*{-8.32cm}
\centerline{\hspace*{7.mm}
\psfig{figure=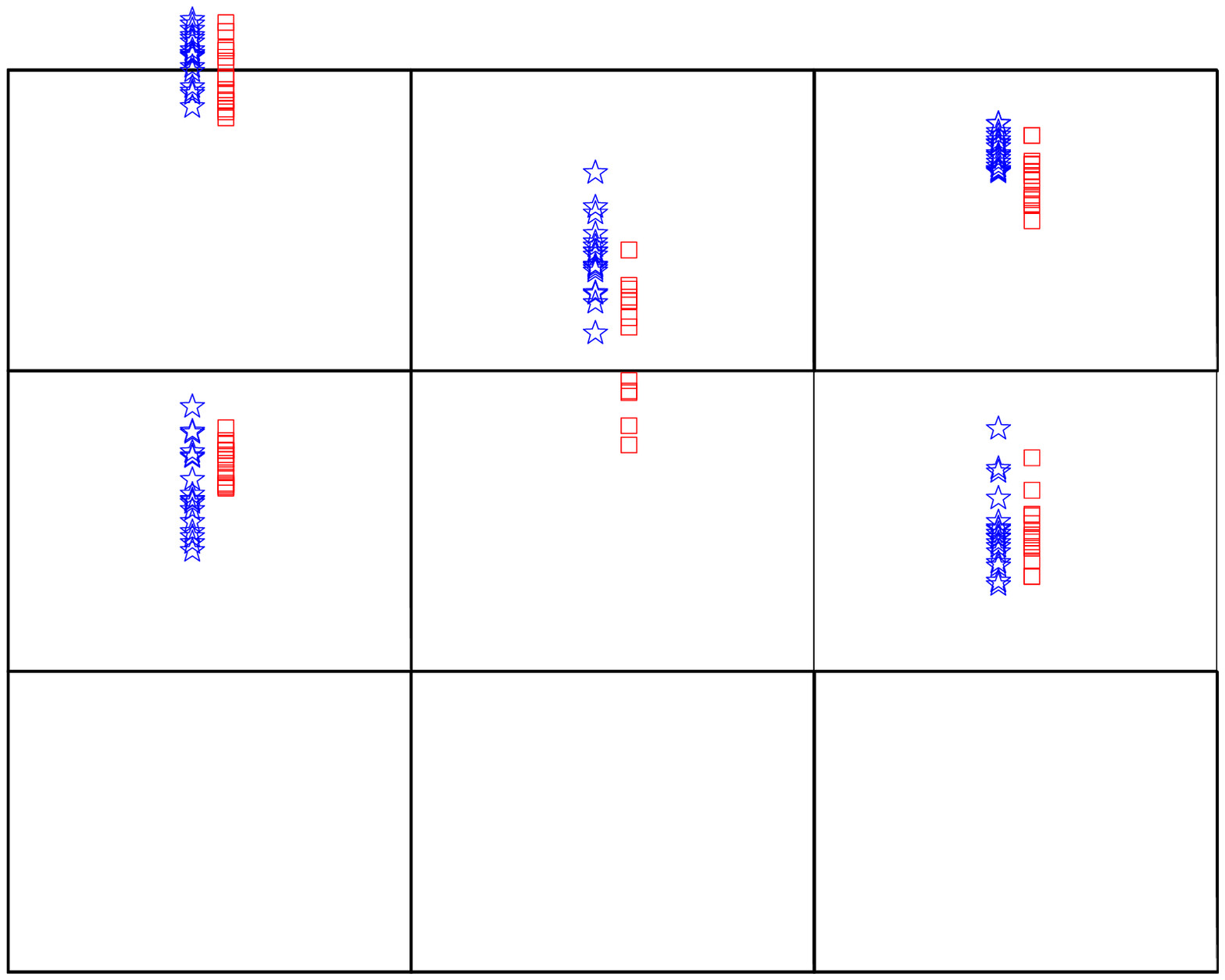,height=7.67cm,width=9.59cm,angle=-90.}
}
\vspace*{5mm}
\caption[]{\baselineskip=0.9\normalbaselineskip
Reddening-corrected line ratios relative to \Ha\ for
ellipticals in the ``RSA sample'', superposed on Fig.~3 of Binette \etal
(1994). The curves represent models of photoionization by old post-AGB
stars, for different gas abundances and burst ages (see bottom center
panel), as a function of the ionization parameter $U$. Symbols: open
starred pentagons represent data for the nuclear 3$''$ of gassy
ellipticals in the ``RSA sample''; open squares represent data for
the extended emission-line region of the same ellipticals (the data
are plotted at arbitrary values of $U$). 
}
\label{f:BMSBplot}
\end{figure}

\paragraph{Emission-line intensity measurements} $\!\!\!\!$were
performed using Gaussian fits to the emission lines in the
dereddened pure emission spectra. The detected lines typically
comprised \OII\,(\lda3727), \NeIII\,(\lda3869), \Hb\,(\lda4861),
\OIII\,(\lda\lda4959, 5007), \OI\,(\lda\lda6300,\,6363), \Ha\,(\lda6563),
\NII\,(\lda\lda6548,\,6583), and \SII\,(\lda\lda6716, 6731). 
The emission-line spectra were then classified using diagnostic
diagrams involving reddening-corrected line intensity ratios (Veilleux
\& Osterbrock 1987). Any residual correction for reddening intrinsic
to the line-emitting regions was performed using the observed \Ha/\Hb\
line ratio, assuming case B recombination, and the Galactic reddening
curve. The results are depicted in Figs.~\ref{f:VO87plots} and
\ref{f:BMSBplot}.  
\subsection{Possible Ionization Mechanisms}

A glance at Fig.~\ref{f:VO87plots} reveals that the ionized gas in
the ``normal'' ellipticals in this sample is (virtually)
always of the LINER class (see also Section 2.2.1). 
Several potential ionization mechanisms have been proposed to account for 
LINER-type spectra, including photoionization by a dilute non-stellar
power-law continuum, shock heating, and turbulent mixing layers (see
Filippenko 1996 for a recent review). In the case of LINERs with
spatially compact emission, Ho, Filippenko \& Sargent (1993) found
that photoionization by a power-law continuum provides the best
overall fit to the observed line ratios. However, the radial variation
of the line ratios for ellipticals with extended emission (cf.\
Fig.~\ref{f:VO87plots}) does not seem consistent with a nuclear source of
ionization: The {\it ionization parameter\/} $U$ (the ratio of the ionizing
photon density to the nucleon density at the face of the gas cloud) of
the gas in the outer regions seems to be consistent with that of the
nuclear gas (cf.\ the model curves 
in Fig.~\ref{f:VO87plots}). If the ionizing agent were a nuclear
source, the gas density would be required to track a $r^{-2}$ form
closely (assuming the covering factor of the gas to be small),
whereas X-ray measurements show that the gas density 
typically varies with radius as $r^{-3/2}$ (e.g., Trinchieri \etal
1997). Taken at face value, this suggests that the gas in these
ellipticals is ionized  by a {\it distributed\/} source.
%of ionization. 
In the remainder of this contribution I will discuss the 
applicability of two such distributed sources: Post-AGB stars and hot
electrons in the X-ray-emitting gas. \\ [-5ex]

\paragraph{Photoionization by Post-AGB Stars} ~ \\ [0.25mm]
As already discussed in Sect.\ 2, this ionization mechanism
provides an obvious explanation for the observed correlation between
emission-line flux and local stellar light flux, and the post-AGB
stars are generally able to produce the observed \Ha+\NII\
luminosities to within a factor of 2. What about the line ratios\,?
Fig.~\ref{f:BMSBplot} depicts a comparison of the measured line ratios 
with photoionization calculations by Binette \etal (1994), for
different gas abundances and ages of the underlying stellar population. 
Fig.~\ref{f:BMSBplot} shows that the measured line ratios are
typically quite well matched by Binette {\it et al.}'s models for $U
\sim 10^{-4}$ except for the \OIII/\Ha\ ratio, which seems to be
somewhat higher than predicted by their models. It should be noted
that their model did not consider other old hot stars such as \eg
AGB-Manqu\'e stars which likely dominate the (stellar) production of
ionizing photons in giant ellipticals (see \eg Brown \etal
1997). Also, the effect of stellar metallicity on the Lyman continuum
output of post-AGB stars still has to be included. Clearly, more work
on the properties of the ionizing radiation from UV-bright stars in
ellipticals will help in appreciating the potential of this ionization
mechanism. 
\\ [-5ex]

\paragraph{Hot Electrons in X-ray-emitting Gas} ~ \\ [0.25mm]
Another ample source of heat energy is available for ellipticals
containing hot gas:\ {\it Thermal conduction\/} by hot electrons transfers
energy from the hot gas to the cooler clouds, by a flux
$F_{\scrm{cond}}$ that saturates
at at rate $\sim 5\,Pv_{\scrm{s}}$, the product of the pressure and the
sound speed (Cowie \& McKee 1977). 
As the flux of photoelectric heat energy into a cloud is 
\[ 
F_{\scrm{ph}} \; \sim \; \frac{Pc\,U(\overline{E}_{\scrm{ph}} - 13.6
\, \mbox{eV})}{2.3\,kT_{\scrm{warm}}} \]
where $\overline{E}_{\scrm{ph}}$ is the mean energy per ionizing
photon and $T_{\scrm{warm}} \sim 10^4$ K the temperature of the ionized
gas, the ratio of the two energy fluxes is 
\[ \frac{F_{\scrm{cond}}}{F_{\scrm{ph}}} \sim 7.5 \left( \frac{U}{10^{-4}}
\right)^{-1} \left( \frac{v_{\scrm{s}}}{300\;\mbox{km\,s$^{-1}$}}
\right) \left( \frac{\overline{E}_{\scrm{ph}}}{13.6\,\mbox{eV}} - 1
\right)^{-1}\mbox{.} \] 
Note that at the low ionization parameters observed in gassy
ellipticals ($U \sim 10^{-4}$), thermal conduction can easily dominate
photoelectric heating in a hot gas (see also Voit \& Donahue
1997). However, a coupled photoionization/conduction code is needed to
solve for the line emission caused by this excitation mechanism but is not
yet available; in addition, further work will be necessary 
to determine how electrons distribute their energy throughout the cool
clouds as a function of their velocity (see \eg Voit 1991). 

%\section{Concluding Remarks}
%
%Bla bla

\acknowledgments

It is a pleasure to thank the organizing committee of this conference
for a job very well done. It was an honor to be invited to participate
in such an outstanding meeting. This paper is partially based on
observations collected at the European Southern Observatory, La Silla,
Chile. The Isaac Newton Telescope is operated at the island of La
Palma by the Royal Greenwich Observatory in the Spanish Observatorio
del Roque de los Muchachos of the Instituto de Astrofisica de Canarias. 

\begin{question}{G.\ Trinchieri}
I'd like to comment that while {\sl AXAF\/} will of course be great
for this kind of work, also {\sl ROSAT\/} HRI data has already shown that
there is a remarkable 
correlation between the morphology of the X-ray emission and the
\Ha+\NII\ emission in a few objects. One is in fact NGC~5846, the same
galaxy for which you showed that the dust correlates beautifully with
the \Ha+\NII\ emission. The {\sl ROSAT\/} PSPC data for that galaxy even
suggest (the spatial resolution of PSPC data does not allow a stronger
statement here) that the feature pointing to the NE in both X-ray--
and \Ha\ emission is cooler than the surrounding gas at the same
radius. All of this seems to fit nicely into the ``evaporation flow''
picture. 
\end{question}
\begin{answer}{P.\ Goudfrooij}
Thanks for the comment, sounds good\,!
\end{answer}
\begin{question}{B.\ Rocca-Volmerange}
The metallicity effect can vary the number of ionizing photons. Did
you include this effect in your modelling of the \Ha\ emission? 
\end{question}
\begin{answer}{P.\ Goudfrooij}
No, I didn't. I considered the model of Binette et al.\ (1994) who
used the Sch\"onberger PAGB tracks for solar metallicity. I agree it
would be very useful to re-consider Binette et al.'s model when the
effects of metallicity on PAGB (and AGB-Manqu\'e) evolution can be
evaluated.  
\end{answer} 
\begin{question}{R.\ Bower}
How do the gas properties depend on environment? For instance, if the
gas comes from the accretion of gas-rich satellites, wouldn't you
expect this mechanism to shut off in environments with 
high velocity dispersion? 
\end{question}
\begin{answer}{P.\ Goudfrooij}
Within the sample I presented, there is indeed a hint that gassy
ellipticals are preferentially found in groups or in the field. 
There are however exceptions, \eg M\,84, M\,87 and
NGC~4696 which are in the central regions of clusters. In those cases,
I regard the more likely origin of the gas to be {\it stripping\/} from a 
spiral passing by too closely for its own comfort, instead of accretion
of a gas-rich galaxy as a whole. This would be consistent with
the fact that none of these gassy cluster ellipticals show any
signs of shells. 
\end{answer}
\begin{question}{R.\ Pogge}
A comment about your use of the Veilleux \& Osterbrock (1987)
diagnostic diagrams. I would be reluctant to read too much into the
various model curves they plot as regards your line data for the
ellipticals. These models assume single clouds in which all of the
electrons responsible for collisionally exciting the forbidden lines
are photoelectrons from Hydrogen. It is pretty clear from the
extended X-ray emission and \Ha\ correlations that a a lot
more is going on these systems, so all bets are off. I think your point
about needing to consider composite conduction \& photoionization
models is a more promising path to follow. There are a number of other
ways to heat gas in these systems, for example turbulent mixing layers
like those considered by Binette and collaborators. 
\end{question}
\begin{answer}{P.\ Goudfrooij}
Sure Rick, I agree with your comments. As to the simple models
depicted in the Veilleux \& Osterbrock diagrams:\ I merely mentioned
them to show that the outer emission regions typically do not have a
lower ionization parameter than the nuclear emission, \ie that the
source of ionizing photons seems to have an extended distribution. As
to models involving turbulent mixing layers, these typically give rise
to the ``type II'' emission-line ratios seen in central cluster
galaxies featuring ``cooling-flows'' (cf.\ Heckman \etal 1989),
whereas ``my'' sample of  
giant ellipticals typically show ``type I'' line ratios.  
\end{answer}
%\begin{question}{V.\ Avila-Reese}
%Which are the most accepted stellar-to-hot gas ratios in
%early-type galaxies?
%\end{question}
%\begin{answer}{P.\ Goudfrooij}
%
%\end{answer}


\begin{references}
\baselineskip=0.85\normalbaselineskip
\small
\reference Baum, S.\,A., Heckman, T.\,M., 1989, \apj\ 336, 702 
\reference Binette, L., Magris, C.\,G., Stasi\'nska, G., Bruzual,
  A.\,G., 1994, \aap\ 292, 13
%\reference Braine, J., Henkel, C., Wiklind, T., 1997, \aap\ 321, 765 
\reference Bregman, J.\,N., Hogg, D.\,E., Roberts, M.\,S., 1992, \apj\
  387, 484  
\reference Brinks, E., 1990, in: {\it The Interstellar Medium in
  Galaxies}, eds.\ H.\,A.\ Thronson \& J.\,M.\ Shull, Kluwer,
  Dordrecht, p.\ 39 
\reference Brown, T.\,M., Ferguson, H.\,C., Davidsen, A.\,F., Dorman, B.,
  1997, \apj\ 482, 685 
\reference Buson, L.\,M., Sadler, E.\,M., Zeilinger, W.\,W., et al.,
  1993, \aap\ 280, 409
\reference Caldwell, N., 1984, \pasp\ 96, 287
%\reference Canizares, C.\,R., Fabbiano, G., Trinchieri, G., 1987,
%  \apj\ 312, 503 
\reference Cowie L.\,L., McKee C.\,F., 1977, \apj\ 211, 135
%\reference Davies, R.\,L., Sadler, E.\,M., Peletier, R.\,F., 1993,
%  \mnras\ 262, 650  
\reference de Jong, T., N{\o}rgaard-Nielsen, H.\,U., Hansen, L.,
  J{\o}r\-gen\-sen, H.\,E., 1990, \aap\ 232, 317
\reference Donahue, M., Voit, G.\,M., 1997, \apj\ 486, 242
\reference Eskridge, P.\,B., Fabbiano, G., Kim D.-W., 1995, \apjs\ 97, 141
\reference Faber, S.\,M., Gallagher, J.\,S., 1976, \apj\ 204, 365
\reference Faber, S.\,M., Wegner, G., Burstein, D., et al., 1989, \apjs\
  69, 763  
\reference Fabian, A.\,C., Nulsen, P.\,E.\,J., Canizares, C.\,R., 1991,
  \aapr\ 2, 191
\reference Fabian, A.\,C., Canizares, C.\,R., B\"ohringer, H., 1994,
  \apj\ 425, 40
\reference Ferland, G.\,J., Netzer, H., 1983, \apj\ 264, 105
%\reference Forbes, D.\,A., 1991, \mnras\ 249, 779 
\reference Forman, W.\,R., Jones, C., Tucker, W., 1985, \apj\ 293, 102
\reference Filippenko, A.\,V. 1996, in: {\it The
  Physics of LINERs in View of Recent Observations}, eds.\ M.\
  Eracleous et al., ASP, San Francisco, p.\ 17
\reference Goudfrooij, P., 1994, Ph.\,D.\ thesis, University of
  Amsterdam, The Netherlands
\reference Goudfrooij, P., 1997, in: {\it The Nature of Elliptical
  Galaxies}, eds.\ M.\ Arnaboldi, G.\,S.\ da Costa, \& P.\ Saha,
  ASP, San Francisco, p.\ 338
%\reference Goudfrooij, P.\ \etalk 1994a, \aaps\ 104, 179 
\reference Goudfrooij, P., Hansen, L., J{\o}rgensen, 
  H.\,E., N{\o}rgaard-Nielsen, H.\,U., 1994, \aaps\ 105, 341 
%\reference Goudfrooij, P.\ \etalk 1994c, \mnras\ 271, 833
%\reference Goudfrooij, P., de Jong, T., 1995, \aap\ 298, 784 
\reference Goudfrooij, P., Trinchieri, G., 1998, \aap\ 330, 123
\reference Haynes, M.\,P., Giovanelli, R., 1984, \aj\ 89, 758 
\reference Heckman, T.\,M., 1980, \aap\ 87, 152 
\reference Heckman, T.\,M., Baum, S.\,A., van Breugel, W.\,J.\,M.,
  McCarthy, P., 1989, \apj\ 338, 48 
\reference Ho, L.\,C., Filippenko, A.\,V., Sargent, W.\,L.\,W., 1993,
  \apj\ 417, 63
\reference Hu, E.\,M., Cowie, L.\,L., Wang, Z., 1985, \apjs\ 59, 447 
\reference Humason, M.\,L., Mayall, N.\,U., Sandage, A., 1956, \aj\ 61,
  97
\reference Kim, D.-W., 1989, \apj\ 346, 653
\reference Kim, D.-W., Fabbiano, G., Trinchieri, G., 1992, \apj\ 393,
  134 
\reference Knapp, G.\,R., 1990, in: {\it The Interstellar Medium in
  Galaxies}, eds.\ H.\,A.\ Thronson \& J.\,M.\ Shull, Kluwer,
  Dordrecht, p.\ 3
\reference Knapp, G.\,R., Turner, E.\,L., Cunniffe, P.\,E., 1985, \aj\
  90, 454
%\reference Knapp, G.\,R., Guharthakurta, P., Kim, D.-W., Jura, M.,
%  1989, \apjs\ 70, 329 
\reference Lees, J.\,F., Knapp, G\,R., Rupen, M.\,P., Phillips,
  T.\,G., 1991, \apj\ 379, 177
\reference Macchetto, F., Pastoriza, M., Caon, N., et al., 1996, \aaps\
  120, 463
\reference Mathews, W.\,G., Baker, J.\,C., 1971, \apj\ 170, 241
\reference Mayall, N.\,U., 1958, in: {\it Comparison of the
  Large-Scale Structure of the Galactic System with that of other
  Systems}, ed.\  N.\,G.\ Roman, Cambridge University Press
%, Cambridge
\reference Mihalas, D., Binney, J.\,J., 1981, {\it Galactic
  Astronomy}, Freeman, San Francisco 
\reference Ostriker, J.\,P., 1990, in: {\it The Interstellar Medium in
  Galaxies}, eds.\ H.\,A.\ Thronson \& J.\,M.\ Shull, Kluwer,
  Dordrecht, p.\ 543
\reference Phillips, M.\,M., Jenkins, C.\,R., Dopita, M.\,A., Sadler,
  E.\,M., Binette, L., 1986, \aj\ 91, 1062
\reference Roberts, M.\,S., Hogg, D.\,E., Bregman, J.\,E., Forman, W.\,R.,
  Jones, C., 1991, \apjs\ 75, 751 
%\reference Rowan-Robinson, M., 1990, in: {\it The Interstellar Medium in
%  Galaxies}, eds.\ H.\,A.\ Thronson \& J.\,M.\ Shull, Kluwer,
%  Dordrecht, p.\ 121. 
\reference Shields, J.\,C., 1991, \aj\ 102, 1314
\reference Singh, K.\,P., Bhat, P.\,N., Prabhu, T.\,P., Kembhavi, A.\,K.,
  1995, \aap\ 302, 658 
%\reference Sparks, W.\,B., 1992, \apj\ 393, 66
\reference Sparks, W.\,B., Macchetto, F., Golombek, D., 1989, \apj\
  345, 513
%\reference Sparks, W.\,B., Jedrzejewski, R.\,I., Macchetto, F., 1994, in:
%  ``The soft X-ray cosmos'', eds.\ E.\ Schlegel \& R.\ Petre, AIP
%  Press, New York, p.\ 389 
\reference Tielens, A.\,G.\,G.\,M., 1990, in: {\it Carbon in the Galaxy:
  Studies from Earth and Space}, eds.\ J.\ Tarter et al., NASA Conf.\
  Proc.\ No.\ 3063, Washington, D.\,C., p.\ 59
\reference Tielens, A.\,G.\,G.\,M., McKee, C.\,F., Seab, C.\,G.,
  Hollenbach, D.\,J., 1994, \apj\ 431, 321 
\reference Trinchieri, G., di Serego Alighieri, S., 1991, \aj\ 101, 1647 
\reference Trinchieri, G., Noris, L., di Serego Alighieri, S., 1997,
  \aap\ 326, 565
\reference Tinsley, B.\,M., 1980, {\it Fundamentals of Cosmic Physics\/} 5,
  287 
\reference Van Dokkum, P.\,G., Franx, M., 1995, \aj\ 110, 2027 
\reference Voit, G.\,M., 1991, \apj\ 377, 158
%\reference V\'{e}ron-Cetty, M.\,P., V\'{e}ron, P., 1988, \aap\ 204,
  28
%\reference White, R.\,E.\ III, Chevalier, R.\,A., 1983, \apj\ 275, 69 
%\reference Wiklind, T., Combes, F., Henkel, C., 1995, \aap\ 297, 643 
%\reference Young, J.\,S., 1990, in: {\it The Interstellar Medium in
%  Galaxies}, eds.\ H.\,A.\ Thronson \& J.\,M.\ Shull, Kluwer,
%  Dordrecht, p.\ 67 
\end{references}
\end{document}